\documentclass[twocolumn]{revtex4}
\usepackage{graphicx,color}
\usepackage{epstopdf}
\usepackage{amssymb}
 \begin{document}
\title{Correlations, fluctuations and stability of a finite-size network of coupled oscillators}

\author{Michael A. Buice}
\author{Carson C. Chow}
\affiliation{Laboratory of Biological  Modeling, NIDDK, NIH, Bethesda, MD}
\date{\today}

\begin{abstract}
The incoherent state of the Kuramoto model of coupled oscillators exhibits marginal modes in mean field theory.  We demonstrate that corrections due to finite size effects render these modes stable in the subcritical case, i.e.~when the population is not synchronous.  This demonstration is facilitated by the construction of a non-equilibrium statistical field theoretic formulation of a generic model of coupled oscillators. This theory is consistent with previous results.   In the all-to-all case, the fluctuations in this theory are due completely to finite size corrections, which can be calculated in an expansion in $1/N$, where $N$ is the number of oscillators.  The $N\rightarrow \infty$ limit of this theory is what is traditionally called mean field theory for the Kuramoto model.
\end{abstract} 

\maketitle

\section{Introduction}


Systems of coupled oscillators have been used to describe  the dynamics of an
extraordinary range of phenomena \cite{winfree}, including networks of
neurons \cite{liu,golomb}, synchronization of blinking fireflies
\cite{ermentrout84,ermentrout91}, chorusing of chirping crickets \cite{walker},  neutrino flavor oscillations \cite{pantaleone},
arrays of 
lasers \cite{yu}, and coupled Josephson junctions \cite{wies}. 
A common model of coupled oscillators is the Kuramoto model \cite{kuramoto}, which  describes the evolution of $N$ coupled oscillators.  A generalized form is given by
\begin{equation}
	\dot{\theta}_i = \omega_i + \frac{K}{N} \sum_j f(\theta_j - \theta_i)
	\label{eq:kuramoto}
\end{equation}
where $i$ labels the oscillators, $\theta_i$ is the phase of oscillator $i$, $f(\theta)$ is the phase dependent coupling, and the intrinsic driving frequencies $\omega_i$ are distributed according to some distribution $g(\omega)$.  In the original Kuramoto  model,  $f(\theta)=\sin(\theta)$.  Here,  we consider $f$ to be any smooth odd function.  
The system can be characterized by the complex order parameter
\begin{equation}
Z(t) = \sum_j e^{i\theta_j(t)}\equiv r(t) e^{i\Psi(t)}
\label{eq:orderp}
\end{equation}
where the magnitude  $r$ gives a measure of synchrony in the system.

In the  limit of an infinite oscillator system,  Kuramoto showed that there is a bifurcation or
continuous phase transition as the coupling $K$ is increased beyond some critical value, $K_c$ \cite{kuramoto}.   Below the critical point the steady state solution has $r =0$ (the ``incoherent" state).  Beyond the critical point, a new steady state solution with $r >  0$ emerges.  
Strogatz and Mirollo analyzed the linear stability of the incoherent state of this system using a Fokker-Planck formalism \cite{strogatz}. In the absence of external noise, the system displays marginal modes associated with the driving frequencies of the oscillators.  
However, numerical simulations of  the Kuramoto model for a large but finite number of oscillators show that the oscillators quickly settle into the incoherent state below the critical point.   The paradox of why the marginally stable incoherent state seemed to be an attractor in simulations was partially resolved by Strogatz, Mirollo and Matthews~\cite{strogMiroMatt} who demonstrated (within the context of the $N\rightarrow \infty$ limit) that there was a dephasing effect akin to Landau damping in plasma physics  which brought $r$ to zero  with a time constant that is inversely proportional to the width of the frequency distribution.   Recently, Strogatz and Mirollo have shown that the fully locked state $r=1$ is stable  \cite{miroStrog4} but the partially locked state is again marginally stable  \cite{miroStrog3}.  
Although dephasing can explain how the order parameter can go to zero, the question of whether the incoherent state is truly stable for a finite number of oscillators remains unknown.  Even with dephasing, in the infinite oscillator limit the system still has an infinite memory of the initial state so there may be classes of initial conditions for which the order parameter or the density exhibits oscillations.  

The applicability of the results for the infinite size Kuramoto model to a finite-size network of oscillators is largely unknown.
The intractability of the finite size case suggests a statistical approach to understanding the dynamics.  Accordingly, the infinite oscillator theories  should be the limits of some averaging process for a finite system.  While the behavior of a finite system is expected to converge to the ``infinite" oscillator behavior, for a finite number of oscillators the dynamics of the system will exhibit fluctuations.  
For example, Daido \cite{daido3, daido4} considered his analytical treatments of the Kuramoto model using time averages and he was able to compute an analytical estimate of the variance.  In contrast,  we will pursue ensemble averages over oscillator phases and driving frequencies.  As the Kuramoto dynamics are deterministic, this is equivalent to an average over initial phases and driving frequencies.  Furthermore, the averaging process imparts a distinction between the order parameter $Z$ and its magnitude $r$.  Namely, do we consider $\langle Z \rangle$ or $\langle r \rangle = \langle |Z| \rangle$ to be the order parameter?  This is important as the two are not equal.  In keeping with the density as the proper degree of freedom for the system (as in the infinite oscillator theories mentioned above), we assert that $\langle Z \rangle$ is the natural order parameter, as it is obtained via a linear transformation applied to the density.  


 Recently, Hildebrand et al. \cite{hildebrand} produced a kinetic  theory inspired by  plasma physics to describe the fluctuations within the system.  They produced a Bogoliubov-Born-Green-Kirkwood-Yvon (BBGKY) moment hierarchy and truncation at second order in the hierarchy yielded analytical results for the two point correlation function from which the fluctuations in the order parameter could be computed.  At this order, the system still manifested marginal modes.  Going beyond second order was impractical within the kinetic theory formalism.  Thus,   it remained an open question as to whether going to higher order would show that  finite size fluctuations could stabilize the marginal modes.

Here, we introduce a statistical field theory approach to calculate the moments of the distribution function governing the Kuramoto model.   The formalism  is equivalent to the Doi-Peliti path integral method used to derive statistical field theories for Markov processes, even though our model is fully deterministic \cite{peliti, doi1, doi2, janssen}.   The field theoretic action we derive produces exactly the same BBGKY hierarchy of the kinetic theory approach~\cite{hildebrand}.  The advantages of the field theory approach are  that 1) difficult calculations are easily visualized and performed through Feynman graph analysis, 2) the theory is easily extendable and generalizable (e.g. to local coupling), 3) the field theoretic formalism permits the use of the renormalization group (which will be necessary near the critical point), and 4) in the case of the all-to-all homogeneous coupling of the Kuramoto model proper, the formalism results in an expansion in $1/N$ and  verifies that mean field theory is exact in the $N \rightarrow \infty$ limit.  We will demonstrate that this theory predicts that finite size corrections will stabilize the marginal modes of the infinite oscillator theory.  Readers unfamiliar with the tools of field theory are directed to one of the standard texts \cite{zinnjustin}.  A review of field theory for non-equilibrium dynamics is \cite{tauber}.

In section~\ref{sec:fieldTheory}, we present the derivation of the theory and elaborate on this theory's relationship to the BBGKY hierarchy.  In section~\ref{sec:TreeLevel}, we describe the computation of correlation functions in this theory and, in particular, describe the tree level linear response.  This will connect the present work directly with what was computed using the kinetic theory approach \cite{hildebrand}.    In section~\ref{sec:marginal}, after describing two example perturbations, we calculate the one loop correction to the linear response and demonstrate that the modes which are marginal at mean field level are rendered stable by finite size effects.   In addition, we demonstrate how generalized Landau damping arises quite naturally within our formalism. We compare these results to simulations in section~\ref{sec:simulation}.

\section{Field Theory for the Kuramoto Model}
\label{sec:fieldTheory}
 
The Kuramoto model (\ref{eq:kuramoto}) can be described in terms of a density
 of oscillators in $\theta$, $\omega$ space
\begin{equation}
\eta(\theta,\omega,t)=\frac{1}{N}\sum_{i=1}^{N}
\delta(\theta-\theta_i(t))\delta(\omega-\omega_i)
\label{eq:fulldist}
\end{equation}
that obeys the continuity equation
\begin{eqnarray}
\lefteqn{{\cal C}(\eta)  \equiv}  \nonumber \\
&&\frac{\partial \eta}{\partial t}+\omega \frac{\partial \eta}{\partial
  \theta}  \nonumber \\
&+& K\frac{\partial}{\partial
  \theta}\int_{-\infty}^{\infty}\int_{0}^{2\pi}f(\theta'-\theta) \eta(\theta',\omega',t) \eta(\theta,\omega,t)d\theta'
  d\omega' \nonumber \\ 
  &=&0
\label{eq:cont} 
\end{eqnarray}
Equation (\ref{eq:cont}) remains an exact description of the dynamics  of the Kuramoto model (\ref{eq:kuramoto})~\cite{hildebrand}.  Although equation~(\ref{eq:cont}) has the same form as that used by Strogatz and Mirollo \cite{strogMiro}, it is  fundamentally different because solutions need not be smooth.  Rather, the solutions of Eq.~(\ref{eq:cont}) are treated in the sense of distributions as defined by Eq.~(\ref{eq:fulldist}).  As we will show, imposing smooth solutions is equivalent to mean field theory (the infinite oscillator limit).  Drawing an analogy to the kinetic theory of plasmas, Eq.~(\ref{eq:cont}) is equivalent to the Klimontovich equation while the mean field equation used by Strogatz and Mirollo \cite{strogMiro} is equivalent to the Vlasov equation \cite{ichimaru,nicholson}.

Our goal is to construct a  field theory 
 to calculate the response functions and moments of the density $\eta$.  Eventually we will construct a theory akin to a Doi-Peliti field theory~\cite{peliti, doi1, doi2, janssen}, a standard approach for reaction-diffusion systems.  We will arrive at this through a construction using a Martin-Siggia-Rose response field \cite{martin}. Since the model is deterministic, the time evolution of $\eta(\theta, \omega, t)$ serves to map the initial distribution forward in time.  We can therefore represent the functional probability measure ${\cal P}[\eta(\theta, \omega, t)]$ for  the density $\eta(\theta, \omega, t)$ as a delta functional which enforces the deterministic evolution from equation~(\ref{eq:cont}) along with an expectation taken over the distribution ${\cal P}_0[\eta_0]$ of the initial configuration $\eta_0(\theta, \omega) = \eta(\theta, \omega, t_0)$.  We emphasize that no external  noise is added to our system.  Any statistical uncertainty completely derives  from the distribution of the initial state.  
 Hence we arrive at the following path integral,
\begin{eqnarray}
\lefteqn{{\cal P}[\eta(\theta, \omega, t)] } \nonumber \\ &=&\int {\cal D}\eta_0{\cal P}_0[\eta_0] \delta  \left [N\left \{ {\cal C}(\eta(\theta,\omega,t))  - \delta(t - t_0) \eta_0(\theta, \omega)\right \} \right ] \nonumber \\
  \label{eq:delta}
\end{eqnarray}
 The definition of the delta functional contains an arbitrary scaling factor, which we have taken to be $N$.  We will show later that this  choice is necessary for the field theory to correctly describe the statistics of $\eta$.
The probability measure obeys the normalization condition
$1 = \int {\cal D} \eta {\cal P}[\eta]$.

We first write the generalized Fourier decomposition of the  delta functional.
\begin{eqnarray}
	\lefteqn{ {\cal P}[\eta(\theta, \omega, t)] } \nonumber \\
	&=&  \int {\cal D}\tilde{\eta}{\cal D}\eta_0{\cal P}_0[\eta_0]   \nonumber \\
	&\times& \exp\left (-N\int d\theta d\omega dt \,\tilde{\eta} \left [{\cal C}(\eta)  
     -\delta(t-t_0)  \eta_0(\theta, \omega)\right]\right ) \nonumber \\
\end{eqnarray} 
where $\tilde{\eta}(\theta, \omega, t)$ is usually called the ``response field" after Martin-Siggia-Rose \cite{martin} and the integral is taken along the imaginary axis.  It is more convenient to work with the generalized probability including the response field:
\begin{eqnarray}
	 \tilde{{\cal P}}[\eta, \tilde{\eta}] 
	&=& \int{\cal D}\eta_0{\cal P}_0[\eta_0] \exp\left (-N\int d\theta d\omega dt \,\tilde{\eta} {\cal C}(\eta)\right . \nonumber \\
 &+& \left .N \int d\theta d\omega \tilde{\eta}(\theta, \omega, t_0) \eta_0(\theta, \omega) \right )
  \label{eq:nDist}
\end{eqnarray}
which obeys the normalization
\begin{equation}
	1 = \int {\cal D} \eta{\cal D}\tilde{\eta} \tilde{{\cal P}}[\eta, \tilde{\eta}; \eta_0]
\end{equation}

We now compute the integral over  $\eta_0$ in equation~(\ref{eq:nDist}), which is an ensemble average over initial phases and driving frequencies.  We assume that the initial phases and driving frequencies for each of the $N$ oscillators are independent and obey the distribution $\rho_0(\theta, \omega)$.    ${\cal P}[\eta_0]$ represents the functional probability distribution for the initial number density of oscillators. Noting that $\eta_0(\theta, \omega)$ is given by
\begin{equation}
	\eta_0(\theta, \omega) = \frac{1}{N} \sum_i \delta(\theta - \theta_i(0)) \delta(\omega - \omega_i)
\end{equation}
and $\int {\cal D}\eta_0 {\cal P}_0[\eta_0] = \int \prod_i d\theta_i d\omega_i\rho_0(\theta_i,\omega_i)$, one can show that the distribution from equation~(\ref{eq:nDist}) is given by
\begin{eqnarray}
 \tilde{{\cal P}}[\eta, \tilde{\eta}] 	 &=& \exp\left (-N\int d\theta d\omega dt \,\tilde{\eta} {\cal C}(\eta) \right . \nonumber \\
 &+& \left .  N\ln \left \{ 1 + \int d\theta d\omega \left [ e ^{\tilde{\eta}(\theta, \omega, t_0)} - 1 \right ] \rho_0(\theta, \omega) \right \} \right ) \nonumber \\
	  \label{avgdist}
\end{eqnarray}
In deriving Eq.~(\ref{avgdist}), we have used the fact that
\begin{eqnarray}
	&& \int{\cal D}\eta_0{\cal P}_0[\eta_0] \exp\left (N \int d\theta d\omega \tilde{\eta}(\theta, \omega, t_0) \eta_0(\theta, \omega) \right ) \nonumber \\ &=&\int \prod_i d\theta_i d\omega_i \rho_0(\theta_i, \omega_i) \exp\left (\sum_i \tilde{\eta}(\theta_i, \omega_i, t_0)  \right )   \nonumber \\
	&=&  \left ( \int d\theta d\omega \rho_0(\theta, \omega)e^{\tilde{\eta}(\theta, \omega, t_0)} \right )^N \nonumber \\
 &=& \exp\left(N \ln \left [1+ \int d\theta d\omega \rho_0(\theta, \omega)\left ( e^{\tilde{\eta}(\theta, \omega,t_0)} -1 \right )  \right ] \right)\nonumber\\
	\label{eq:icGenFunc}
\end{eqnarray}
 We see that  the fluctuations (i.e.~terms non-linear in $\tilde{\eta}$) appear only in the initial condition of (\ref{avgdist}), which is to be expected since the Kuramoto system is deterministic.  In this form the continuity equation (\ref{eq:cont}) appears as a Langevin equation sourced by the noise from the initial state.

Although the noise is contained entirely within the initial conditions, it is still relatively complicated.  We can simplify the structure of the noise in (\ref{avgdist}) by performing the following transformation \cite{janssen}:
\begin{eqnarray}
	\varphi(\theta, \omega, t) &=& \eta\exp(-\tilde{\eta}) \nonumber \\
	\tilde{\varphi}(\theta, \omega, t) +1 &=& \exp(\tilde{\eta}) 
	\label{eq:doipeliti}
\end{eqnarray}
%
Under the transformation (\ref{eq:doipeliti}), $\tilde{\cal P}[n,\tilde{n}]$ becomes $\tilde{\cal P}[\varphi, \tilde{\varphi}]$, which is given by
\begin{equation}
	\tilde{{\cal P}}[\varphi, \tilde{\varphi}] = \exp \left ( -N S[\varphi, \tilde{\varphi} ] \right )
\label{eq:dist}
\end{equation}
where the {\em action} $S[\varphi, \tilde{\varphi} ]$ is 
\begin{eqnarray}
	S[\varphi, \tilde{\varphi}] &=& \int d\omega d\theta dt \left [   \tilde{\varphi}\left (\frac{\partial}{\partial t} + \omega \frac{\partial}{\partial \theta} \right ) \varphi  \right . \nonumber \\ 
	&+& \left .  K\int d\omega' d\theta' \left (\tilde{\varphi}' \tilde{\varphi} + \tilde{\varphi}\right) \frac{\partial}{\partial \theta}\left \{ f(\theta' - \theta) \varphi' \varphi \right \} \right ] \nonumber \\ 
	&-&\ln \left [ 1 + \int d\theta d\omega  \tilde{\varphi}(\theta, \omega, t_0) \rho_0(\theta, \omega)\right ]
\label{eq:action1}
\end{eqnarray}
The form (\ref{eq:action1}) is obtained from the transformation (\ref{eq:doipeliti}) only after several integrations by parts.  In most cases, these integrations do not yield boundary terms because of the periodic domain (i.e. those in $\theta$).  In the case of the $\partial_t$ operator, however, we are left with boundary terms of the form $[ \ln (\tilde{\varphi} + 1) -1 ]\tilde{\varphi}\varphi$.  These terms will not affect computations of the moments because of the causality of the propagator (see section~\ref{sec:prop}). 


We are interested in fluctuations around a smooth solution $\rho(\theta, \omega, t)$ of the continuity equation~(\ref{eq:cont}) with initial condition $\rho(\theta, \omega, t_0) = \rho_0(\theta, \omega)$.
 We transform
 the field variables via $\psi = \varphi - \rho$ and $\tilde{\psi} = \tilde{\varphi}$ in (\ref{eq:action1}) and obtain the following action:
  \begin{eqnarray}
	S[\psi, \tilde{\psi}] = \int d\omega d\theta dt \tilde{\psi}\left [   \left (\frac{\partial}{\partial t} + \omega \frac{\partial}{\partial \theta} \right ) \psi \right . &&\nonumber \\
	+ \left . K\int d\omega' d\theta'  \frac{\partial}{\partial \theta}\left \{ f(\theta' - \theta)\left (\psi' \rho +\rho' \psi + \psi' \psi  \right ) \right \}
	\right . &&\nonumber \\
	+ \left . K\int d\omega' d\theta' \tilde{\psi}'  \frac{\partial}{\partial \theta}\left \{ f(\theta' - \theta)\left ( \psi'  + \rho' \right ) \left (\psi + \rho \right ) \right \} \right ] &&\nonumber \\
	- \sum_{k=2} ^\infty\frac{(-1)^{k+1}}{k} \left [ \int d\theta d\omega  \tilde{\psi}(\theta, \omega, t_0) \rho_0(\theta, \omega)\right ]^k&&
	\label{eq:action}
\end{eqnarray}
For fluctuations about the incoherent state: $\rho(\theta, \omega, t) = \rho_0(\theta,\omega)=g(\omega)/2\pi$, where $g(\omega)$ is a fixed frequency distribution.  The incoherent state is an exact solution of the continuity equation (\ref{eq:cont}).  Due to the homogeneity in $\theta$ and the derivative couplings, there are no corrections to it at any order in $1/N$.  
The action (\ref{eq:action}) with $\rho=g(\omega)/2\pi$ therefore describes fluctuations about the true mean distribution of the theory, i.e.~$\langle \eta(\theta, \omega, t)\rangle = g(\omega)/2\pi$.
We can evaluate  the moments of the probability distribution (\ref{eq:dist}) with (\ref{eq:action}) using the method of steepest descents, which treats $1/N$ as an expansion parameter.  This is a standard method in field theory which produces the loop expansion \cite{zinnjustin}.  
We first separate the action (\ref{eq:action}) into ``free" and ``interacting" terms.
\begin{equation}
	S[\psi, \tilde{\psi}] = S_F[\psi, \tilde{\psi}] + S_I[\psi, \tilde{\psi}]
\end{equation}
where
\begin{eqnarray}
	\lefteqn{S_F[\psi, \tilde{\psi}]} \nonumber \\ &=& \int d\omega d\theta dt \tilde{\psi}\left [   \left (\frac{\partial}{\partial t} + \omega \frac{\partial}{\partial \theta} \right ) \psi \right . \nonumber \\
	&+& \left .  K\int d\omega' d\theta'  \frac{\partial}{\partial \theta}\left \{ f(\theta' - \theta)\left (\psi' \rho +\rho' \psi   \right ) \right \}
	\right ] \nonumber \\
	&\equiv&  \int dx dt  dx' dt' \tilde{\psi}(x',t')\Gamma_0(x, t; x', t')\psi(x,t) \nonumber \\ \label{eq:freeAction} 
\end{eqnarray}
\begin{eqnarray}
	\lefteqn{S_I[\psi, \tilde{\psi}] } \nonumber \\ &=&  \int d\omega d\theta dt \tilde{\psi}\left [   K\int d\omega' d\theta'  \frac{\partial}{\partial \theta}\left \{ f(\theta' - \theta)\left ( \psi' \psi  \right ) \right \}
	\right . \nonumber \\
	&+& \left . K\int d\omega' d\theta' \tilde{\psi}'  \frac{\partial}{\partial \theta}\left \{ f(\theta' - \theta)\left ( \psi'  + \rho' \right ) \left (\psi + \rho \right ) \right \} \right ] \nonumber \\
	&-& \sum_{k=2} ^\infty\frac{(-1)^{k+1}}{k} \left [ \int d\theta d\omega  \tilde{\psi}(\theta, \omega, t_0) \rho_0(\theta, \omega)\right ]^k \label{eq:InteractingAction}
\end{eqnarray}
 In deriving the loop expansion, the action is expanded around a saddle point, resulting in an asymptotic series whose terms consist of moments of the Gaussian functional defined by the  terms in the action~(\ref{eq:action}) which are bilinear in $\psi$ and $\tilde{\psi}$, i.e. $S_F[\psi, \tilde{\psi}]$.  Hence, the loop expansion terms consist of various combinations of the inverse of the operator $\Gamma_0$, defined by $S_F$, called the bare propagator, with the higher order terms in the action, called the vertices.  Vertices are given by the terms in $S_I$.

\begin{figure}
	\scalebox{0.25}{\includegraphics{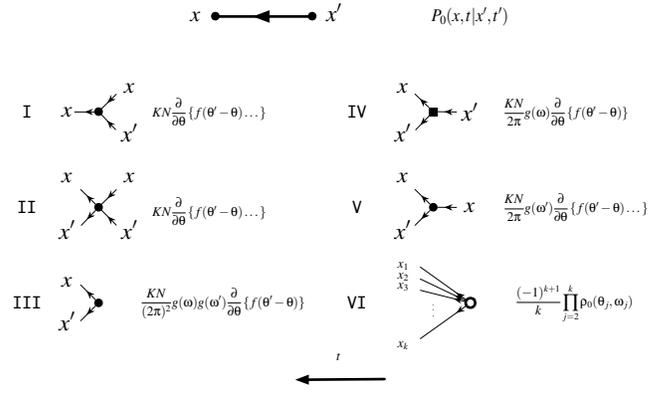}}
	\caption{Diagrammatic (Feynman) rules for the fluctuations about the mean.  Time moves from right to left, as indicated by the arrow.  The bare propagator $P_0(x,t|x',t')$ (see Eq.~(\ref{eq:TreeProp})) connects points at $x'$ to $x$, where $x\equiv\{\theta,\omega\}$.  Each branch of a vertex is labeled by $x$ and $x'$ and is connected to a factor of the propagator at  $x$ or $x'$.  Each vertex  represents an operator given to the right of that vertex.   The ``\dots" on which the derivatives act only include the incoming propagators, but not the outgoing ones.  There are integrations over $\theta, \theta', \omega, \omega'$ and $t$ at each vertex.}
	\label{fig:feynman}
\end{figure}

The terms in  the loop expansion are conveniently represented by diagrams. 
The bare propagator is represented diagrammatically by a line, and should be compared to the variance of a gaussian distribution.  Each term in the action (other than the bilinear term) with $n$ powers of $\psi$  and $m$ powers of $\tilde{\psi}$ is  represented by a vertex with $n$ incoming lines and $m$ outgoing lines.   The initial state vertices produce only outgoing lines and, like the non-initial state or  ``bulk" vertices, are integrated over $\theta, \omega,$ and $t$ for each point at which the operators are defined.  The bulk vertices are represented  by a solid black dot (or square, see Figure~\ref{fig:feynman}) and initial state vertices by an open circle.  The bare propagator and vertices are shown in Figure~\ref{fig:feynman}.  Unlike conventional Feynman diagrams used in field theory, the vertices in Figure~\ref{fig:feynman} represent nonlocal operators defined at multiple points.  In particular, the initial state terms involve operators at a different point for each outgoing line.  Although unconventional, this is the natural way of characterizing the $1/N$ expansion.

Adopting the shorthand notation of $x\equiv\{\theta,\omega\}$,
each arm of a vertex must be connected to a line (propagator) at either $x$ or $x'$ and lines
connect outgoing arms in one vertex to incoming arms in another.  The moment with $n$ powers of $\psi$ and $m$ powers of $\tilde{\psi}$ is calculated by summing all diagrams with
$n$ outgoing lines and $m$ incoming lines.      This means that each diagram will stand for several terms which are equivalent by permutations of the vertices or the edges in the graph, equivalently permutations of the factors of $\tilde{\psi}$ and $\psi$ in the terms in the series expansion.  In a typical field theory, this  results in combinatoric factors.   In the present case, diagrams which are  not topologically distinct can produce different contributions to a given moment.  Nonetheless, we will designate the sum of these terms with a single graph.  Generically, the combinatoric factors we expect are due to the exchange of equivalent vertices, which typically cancel the factorial in the series expansion.  Additionally, each line in a diagram contributes a factor of $1/N$ and each vertex contributes a factor of $N$.  Hence each loop in a diagram carries a factor of $1/N$.  The terms in the expansion without loops are called ``tree level".  The bare propagator is the tree level expansion of $\langle \psi(\theta, \omega, t) \tilde{\psi}(\theta', \omega', t') \rangle$. The tree level expansion of each moment beyond the first carries an additional factor of $1/N$, i.e. the propagator and two-point correlators are each $O(1/N)$.

\emph{Mean field theory} is defined as  the $N \rightarrow \infty$ limit of this field theory.  In the infinite size limit,  all moments higher than the first are zero (provided the terms in the series are not at a singular point, i.e. the onset of synchrony).  Hence, the only surviving terms in the action~(\ref{eq:action}) are those which contribute to the mean of the field at tree level.  These terms sum to give solutions to the continuity equation~(\ref{eq:cont}).  If the initial conditions are smooth, then mean field theory is given by the relevant smooth solution of (\ref{eq:cont}).   In most of the previous work (e.g. \cite{strogMiro, strogMiroMatt}), smooth solutions to (\ref{eq:cont}) were taken as the starting point and hence automatically assumed mean field theory.  

We can now validate our choice of $N$ as the correct scaling factor in the delta functional of (\ref{eq:delta}) by considering the equal time two-point correlator.
Using the definition of $\eta$ from (\ref{eq:fulldist}) we get 
\begin{eqnarray}
\langle \eta(x, t) \eta(x',t) \rangle &=& C(x,x',t) +\rho(x,t) \rho(x',t)  \nonumber \\ 
&-& \frac{1}{N}\rho(x,t) \rho(x',t)  +   \frac{1}{N}\delta(x-x') \rho(x', t) \nonumber \\
\label{eq:2ndmoment}
\end{eqnarray}
where $\rho(x,t)=\langle \eta(x,t)\rangle=\langle \varphi(x,t)\rangle$.
Using the fields $\varphi$ and $\tilde{\varphi}$ (defined in (\ref{eq:doipeliti})) and  taking $\eta$ at different times gives
\begin{eqnarray}
\lefteqn{\langle \eta(x, t) \eta(x',t') \rangle} \nonumber \\  &=& \langle \left [\tilde{\varphi}(x, t)\varphi(x,t)+\varphi(x,t)\right ] \left [\tilde{\varphi}(x', t')\varphi(x',t')+\varphi(x',t')\right ]\rangle \nonumber \\
\label{eq:momentloop}
\end{eqnarray}
for $t>t'$.
The response field has the property that expectation values containing $\tilde{\varphi}(x,t)$ are zero unless another field insertion of $\varphi(x,t)$ is also present but at a later time (this is because the propagator is causal; it is zero for $t - t' \le 0$).  Therefore
\begin{eqnarray}
	\langle \eta(x, t) \eta(x',t') \rangle &=& \langle \varphi(x,t)\varphi(x',t') \rangle \nonumber \\
	&+& \langle \varphi(x,t) \tilde{\varphi}(x', t') \rangle \langle \varphi(x',t')\rangle
	\label{twotime}
\end{eqnarray}
As we will show later when we discuss the propagator in more detail, we have 
\begin{equation}
	\lim_{t \rightarrow t'} \langle \varphi(x,t) \tilde{\varphi}(x', t') \rangle = \frac{1}{N} \delta(x - x')
\end{equation}
Comparing (\ref{twotime}) in the limit $t\rightarrow t'$ with (\ref{eq:2ndmoment}) allows the immediate identification of
\begin{eqnarray}
\langle \varphi(x, t) \varphi(x',t) \rangle&=&C(x,x',t)+\rho(x,t)\rho(x',t) \nonumber \\
 &-& \frac{1}{N}\rho(x,t) \rho(x',t) 
\label{eq:2ndphi}
\end{eqnarray}
$C(x,x',t)$ is the two oscillator ``connected" correlation or moment function.  This is consistent with (\ref{eq:action}) which gives 
\begin{equation}
	\langle \varphi(x, t_0) \varphi(x',t_0) \rangle=\rho(x,t_0)\rho(x',t_0) - \frac{1}{N}\rho(x,t_0) \rho(x',t_0) 
\end{equation}
as the initial condition.
 Thus, comparing the second moment of $\eta$ using the Doi-Peliti fields  (\ref{eq:momentloop})  with the expression from the direct computation given by (\ref{eq:2ndmoment}) shows that
the factor of $N$ in the delta functional of (\ref{eq:delta}) was necessary to obtain the correct scaling for the moments.

The Doi-Peliti action (\ref{eq:action}) can also be derived by considering an
effective  Markov process on a circular lattice representing the angle $\theta$ where the probability of an oscillator moving to a new point on the lattice is determined by its native driving frequency $\omega$ and the relative phases of the other oscillators  (see Appendix~\ref{app:markov}).  This is the primary reason we refer to the theory as being of Doi-Peliti type.  The continuum limit of this process yields a theory described by the action~(\ref{eq:action}).    The Markov picture provides an intuitive description and underscores the fundamental idea that we have produced a \emph{statistical} theory obeyed by a \emph{deterministic} process.   

Although our formalism is statistical, we emphasize that no approximations have been introduced.  The statistical uncertainty is inherited from averaging over the initial phases and driving frequencies.  This formalism could be applied to a wide variety of deterministic dynamical systems that can be represented by a distributional continuity equation like Eq.~(\ref{eq:cont}).   In general, a solution for the moment generating functional for our action (\ref{eq:action}) is as difficult to obtain as solving the original system.  The advantage of formulating the system as a field theory is that a controlled perturbation expansion with the inverse system size as the small parameter is possible.  


\subsection{Relation to Kinetic Theory and Moment Hierarchies}

The theory defined by the action (\ref{eq:action}) is equivalently expressed as a Born-Bogoliubov-Green-Kirkwood-Yvon (BBGKY) moment hierarchy starting with the continuity equation (\ref{eq:cont}).  To construct the moment hierarchy, one takes expectation values of the continuity equation with products of the density, $\eta(\theta, \omega, t)$.  This results in coupled equations of motion for the various moments of $\eta(\theta, \omega, t)$, where each equation depends upon one higher moment in the hierarchy.  The Dyson-Schwinger equations derivable from (\ref{eq:action}) with  $\langle \tilde{\varphi} \rangle = 0$ are exactly the BBGKY hierarchy derived in Ref.~\cite{hildebrand}.  Thus the kinetic theory  and field theory approaches are entirely equivalent.

The moments of $\eta$ can be computed in the BBGKY hierarchy by truncating at some level.   In Ref.~\cite{hildebrand}, this was done at Gaussian order by assuming that the connected three-point function was zero.  The first two equations of the hierarchy then form a closed system which can by solved.  
The first two equations of the BBGKY hierarchy~\cite{hildebrand} are
\begin{eqnarray}
&&\frac{\partial \rho}{\partial t}+\omega \frac{\partial \rho}{\partial \theta} \nonumber \\
&+&K\frac{\partial}{\partial \theta}\int_{-\infty}^{\infty}\int_{0}^{2\pi}f(\theta'-\theta)\rho(x,t)\rho(x',t)d\theta' d\omega' \nonumber\\
&=&-K\frac{\partial}{\partial \theta}\int_{-\infty}^{\infty}\int_{0}^{2\pi}f(\theta'-\theta)C(x,x',t)d\theta' d\omega' \label{eq:bbgky1}
\end{eqnarray}
where $\rho(x,t)=\langle \eta(x,t)\rangle$, and the connected (equal-time) correlation function
\begin{eqnarray}
	C(x,x', t) &=& \langle \eta(x, t) \eta(x', t) \rangle  - \rho(x,t) \rho(x',t) \nonumber \\
	&+& \frac{1}{N}\rho(x,t) \rho(x',t)  - \frac{1}{N} \delta(x-x')  \rho(x', t) \nonumber \\ \label{eq:connected}
\end{eqnarray}
obeys
\begin{widetext}
\begin{eqnarray}
  &&\left \{\frac{\partial}{\partial t}+\omega_1\frac{\partial}{\partial \theta_1}+\omega_2 \frac{\partial}{\partial \theta_2} 
+K\int_{-\infty}^{\infty}\int_{0}^{2\pi} [\frac{\partial}{\partial \theta_1}f(\theta_3-\theta_1) +\frac{\partial}{\partial \theta_2}f(\theta_3-\theta_2)]
 \rho(x_3,t)d\theta_3 d\omega_3 \right \}C(x_1,x_2,t)\nonumber \\
&+& K\int_{-\infty}^{\infty}\int_{0}^{2\pi} \frac{\partial}{\partial \theta_1}f(\theta_3-\theta_1)\rho(x_1,t)C(x_2,x_3,t)d\theta_3 d\omega_3
+ K\int_{-\infty}^{\infty}\int_{0}^{2\pi} \frac{\partial}{\partial \theta_2}f(\theta_3-\theta_2)\rho(x_2,t)C(x_3,x_1,t)d\theta_3 d\omega_3 \} \nonumber \\
&=&-\frac{K}{N}[\frac{\partial}{\partial \theta_1}f(\theta_2-\theta_1)
  +\frac{\partial}{\partial
    \theta_2}f(\theta_1-\theta_2)]\rho(x_1,t)\rho(x_2,t). \nonumber \\
\label{eq:C_rho1} 
\end{eqnarray}
\end{widetext}

In the field theoretic approach, instead of truncating the BBGKY hierarchy, one instead truncates the loop expansion.  Truncating the moment hierarchy at the $m$th order is equivalent to truncating the loop expansion for the $l$th moment at the   $(m-l)$th order.  Thus the solution to the moment equations (\ref{eq:bbgky1}) and (\ref{eq:C_rho1}) is the one loop expression for the first moment and the  tree level expression for the second moment.
 The advantage of using the action (\ref{eq:action}) is that the terms in the perturbation expansion are given automatically by the relevant diagrams  at any level of the hierarchy.  Ref.~\cite{hildebrand} suggested that a higher order in the hierarchy would be necessary to check whether the mean field marginal modes are stable for finite $N$.  We demonstrate below that the field theory facilitates the calculation of the linearization of equation~(\ref{eq:bbgky1}) to higher order in $1/N$ and show that marginal modes are stabilized by finite size fluctuations. 
  
One can compare this approach to the maximum entropy approach of Rangan and Cai~\cite{cai3} for developing consistent moment closures for such hierarchies.  In the moment hierarchy approach of Ref.~\cite{hildebrand}, moment closure is obtained via the somewhat ad hoc approach of setting the $n$th cumulant to zero.  In contrast, Rangan and Cai maximize the entropy of the distribution subject to certain normalization constraints.  The moment closure is facilitated by constraining higher moments from the hierarchy.  However, one still must solve the resulting equations.  In the loop expansion, moment closure is obtained implicitly via truncating the loop expansion.  The loop expansion approach offers the advantage of providing a natural means for determining when the approximation, thus the implicit closure, breaks down and avoids dealing with the moment hierarchy explicitly.  In fact, Rangan and Cai's procedure has a natural interpretation in field theory, namely the minimization of a generalized effective action in terms of various moments.  The simplest and most common is the effective action in terms of the mean field, which is the generating functional of  one partical irreducible (1PI) graphs \cite{zinnjustin}.  The next level of approximation is a generalized effective action (the ``effective action for composite operators" \cite{cornwall}) in terms of the mean and the two-point function (or functions), which is the generating functional of two particle irreducible (2PI)  graphs.   One can continue in this way.  The equations of motion of these effective actions will produce a closure of the moment hierarchy implicit in the action for the theory.   At tree level these equations will be equivalent to those produced by  Rangan and Cai's maximum entropy approach.  The loop expansion allows for systematic corrections to these equations without explicitly invoking higher equations in the hierarchy.

\section{Tree Level Linear Response, Correlations and Fluctuations}
\label{sec:TreeLevel}

As a first example, we reproduce the calculation of the variation of the order parameter $Z$, which was calculated previously using the BBGKY moment hierarchy \cite{hildebrand}.  
To do so requires the calculation of the tree level linear response or bare propagator and the tree level connected two-oscillator correlation function.


	

\subsection{The Propagator}
	\label{sec:prop}
The  propagator $P(\theta, \omega, t| \theta', \omega', t')$ 
is given by the expectation value
\begin{equation}
P(\theta, \omega, t| \theta', \omega', t') = \langle \varphi(\theta, \omega, t) \tilde{\varphi}(\theta', \omega', t') \rangle
\end{equation}
It is the linear response of (\ref{eq:cont}).
This can be shown by considering a small perturbation  $\delta \rho_0$ to the initial state $\rho$ in the action (\ref{eq:action}).  Expanding to first order then yields:
	\begin{equation}
		\delta \rho(\theta, \omega, t) = N\int d\theta' d\omega' \left < \varphi(\theta, \omega, t) \tilde{\varphi}(\theta', \omega', t' )\right > \delta \rho_0(\theta', \omega')
	\end{equation} 
The tree level linear response or bare propagator $P_0(\theta, \omega, t| \theta', \omega', t')\equiv P_0(x,t; x',t')$ is  the functional inverse of the operator $\Gamma_0$  defined by the free part of the action~(\ref{eq:freeAction}).  
The bare propagator is therefore given by \cite{zinnjustin}
	\begin{eqnarray}
		\Gamma_0 \cdot P_0  &\equiv& \int dx'' dt'' \Gamma_0(x, t; x'', t'') P_0 (x'', t'';  x', t') \nonumber \\
		&=& \frac{1}{N} \delta(x - x') \delta(t - t')
		\label{eq:bilinear}
	\end{eqnarray}
	Using the action~(\ref{eq:action}) with Eq.~(\ref{eq:bilinear}) gives
	\begin{widetext}
		\begin{eqnarray} 
\Gamma_0 \cdot P_0  &\equiv& \left [ \frac{\partial}{\partial t} 
+ \omega \frac{\partial}{\partial \theta} 
+ K\frac{\partial}{\partial \theta}\int_{-\infty}^{\infty}\int_{0}^{2\pi}f(\theta_1-\theta)\rho(x_1,t)d\theta_1 d\omega_1\right ] 
P_0(x,x',t-t') \nonumber \\
&+&K\frac{\partial}{\partial \theta}\int_{-\infty}^{\infty}\int_{0}^{2\pi}f(\theta_1-\theta)\rho(x,t)P_0(x_1,x',t-t')d\theta_1 d\omega_1\nonumber \\
	&=&\frac{1}{N}\delta(\theta - \theta') \delta(\omega - \omega') \delta(t - t')
\label{eq:TreeProp}
\end{eqnarray}
\end{widetext}
Due to  the rotational invariance in $\theta$ of $f(\theta)$, $P(x,t;x',t')\equiv P(\theta-\theta',\omega,\omega',t-t')$.

In the incoherent state, $\rho(\theta, \omega, t) = g(\omega)/2\pi$.  Thus, for  $f(\theta)$ odd,  Eq. (\ref{eq:TreeProp}) becomes
\begin{eqnarray}
&&\left [ \frac{\partial}{\partial t}+\omega \frac{\partial}{\partial \theta} 
\right ] P_0(x,x',t-t')  \nonumber \\
&+& K\frac{g(\omega)}{2\pi}\frac{\partial}{\partial \theta}\int_{-\infty}^{\infty}\int_{0}^{2\pi}f(\theta_1-\theta)P_0(x_1,x',t-t')d\theta_1 d\omega_1 \nonumber \\
	&=&	\frac{1}{N}\delta(\theta - \theta') \delta(\omega - \omega') \delta(t - t') 
\label{eq:IncoherentTreeProp}
\end{eqnarray}
We can invert this equation using Fourier and Laplace transforms.

Taking the Fourier transform of Eq.~(\ref{eq:IncoherentTreeProp}) (with respect to $\theta$ and $\theta'$) yields
\begin{eqnarray}
	&&\left [ \frac{\partial}{\partial t}  + i n \omega \right ] P_0(n, \omega; m,  \omega', t - t') \nonumber \\
	&+& in K g(\omega) f(-n) \int_{-\infty}^{\infty} P_0(n, \omega_1; m,  \omega', t - t') d\omega_1 \nonumber \\
	&=& \frac{1}{2\pi N} \delta(t - t') \delta(\omega - \omega') \delta_{n+m}
\end{eqnarray}
where we use the following convention for the Fourier transform
\begin{eqnarray}
	f(n) &=& \frac{1}{2\pi} \int_0^{2 \pi} f(\theta) e^{-i n \theta} d \theta\nonumber \\
	f(\theta) &=& \sum_n f(n) e^{in\theta}
\end{eqnarray}
Hereon, we will suppress the index $m$ since the propagator must be diagonal (i.e. $P_0(n,m))\propto \delta_{m+n}$).  

We Laplace transform in $\tau = t - t'$ to get
\begin{eqnarray}
		&&\left [s  + i n \omega \right ] \tilde{P}_0(n,\omega,  \omega', s) \nonumber \\
		&+& in K g(\omega) f(-n) \int_{-\infty}^{\infty} \tilde{P}_0(n, \omega_1, \omega', s) d\omega_1 \nonumber \\
	&=& \frac{1}{2\pi N}  \delta(\omega - \omega')
\label{eq:lap}
\end{eqnarray}
using the convention
\begin{eqnarray}
	\tilde{f}(s) = \int_0^{\infty} f(t) e^{-s\tau} d\tau \nonumber \\
	f(\tau) = \frac{1}{2\pi i} \int_{\cal L} \tilde{f}(s)e^{s\tau} ds
\end{eqnarray}
where the contour ${\cal L}$ is to the right of all poles in $\tilde{f}(s)$.

We can solve for $\tilde{P}_0(n, \omega, \omega', s)$ using a self-consistency condition.  Integrate (\ref{eq:lap}) over $\omega$ after dividing by $s + in\omega$ to get
\begin{eqnarray}
	&&\int d\omega \tilde{P}_0(n, \omega, \omega', s)\nonumber \\
	 &+& \int d\omega \frac{i n K g(\omega) f(-n)}{s + i n \omega} \int d\omega_1 \tilde{P}_0(n,\omega_1, \omega', s) \nonumber \\
	 &=& \frac{1}{2\pi N} \frac{1}{s + in\omega'}
\end{eqnarray}	 
which we can solve to obtain
\begin{eqnarray}
	\int d\omega \tilde{P}_0(n,\omega, \omega', s) = \frac{1}{2\pi N} \frac{1}{s + in\omega'} \frac{1}{\Lambda_n(s)}
	\label{eq:propIntOne}
\end{eqnarray}
where
\begin{eqnarray}
	\Lambda_n(s) = 1 + i n K  f(-n)\int d\omega \frac{g(\omega)}{s + i n \omega}
	\label{eq:lambda}
\end{eqnarray}
$\Lambda_n(s)$ is defined for ${\rm Re}(s) \le 0$ via analytic continuation.
In the kinetic theory context of an oscillator density obeying the continuity equation (\ref{eq:cont}), $\Lambda_n(s)$ is analogous to a plasma dielectric function~\cite{hildebrand}.
If we assume that $g(\omega)$ is even and $f(\theta)$ is odd, then there is a single real number $s_n$ such that $\Lambda_n(s_n) = 0$. (Mirollo and Strogatz  proved that there is at most one single, real root of (\ref{eq:lambda}) and that it must satisfy ${\rm Re}(s) \ge 0$ \cite{miroStog2, strogatz}.  In our case, $\Lambda_n(s)$ is defined  for ${\rm Re}(s) < 0$ not by (\ref{eq:lambda}), but rather via analytic continuation.) 
Using (\ref{eq:lambda}) in (\ref{eq:lap}) and solving for $\tilde{P}_0(n,\omega, \omega', s)$ gives
\begin{eqnarray}
\tilde{P}_0(n,\omega, \omega', s) &=&  \frac{1}{2\pi N} \frac{\delta(\omega - \omega')}{s + in\omega} \nonumber \\
&-& \frac{1}{2 \pi N} \frac{inKg(\omega) f(-n)}{\left ( s + i n\omega \right) \left ( s + i n \omega' \right )}  \frac{1}{\Lambda_n(s)} \nonumber \\
\label{eq:proptree}
\end{eqnarray}
Here we identify the spectrum with the zeroes of $\Gamma_0$ or, equivalently, the poles of the propagator, as these will determine the time evolution of perturbations.  Analogous to the analysis of Strogatz and Mirollo \cite{strogMiro} we define the  operator ${\cal O}$ by 
\begin{eqnarray}
	{\cal O}[ {b}_n(\omega,  t)] &\equiv& i n \omega {b}_n(\omega,  t) \nonumber \\
	&+& in K  f_{-n}   \int_{-\infty}^{\infty}b_n(\omega_1,  t) g(\omega_1) d\omega_1
	\label{eq:linearO}
\end{eqnarray}
(cf. equation~(\ref{eq:smlin}) below).
The continuous spectrum of ${\cal O}$ consists of the frequencies $in\omega$ whereas the discrete spectrum (according to Ref.~\cite{strogMiro}) only exists for $K > K_c$.  Consistent with that approach (i.e. linear operator theory), we identify the poles in $P$ due to $s + in\omega$ as the continuous spectrum and those due to the zeroes of $\Lambda_n(s)$ as the discrete spectrum.  If  $\Lambda_n(s)$ is not analytically continued for ${\rm Re}(s) < 0$, it will not have zeroes for that domain, as in Ref.~\cite{strogMiro}.  However, zeros can exist for ${\rm Re}(s) < 0$ when $\Lambda_n(s)$ is analytically continued and this is why analytic continuation is of such crucial importance to the conclusions of Ref.~\cite{strogMiroMatt}.


\subsection{Correlation Function}
\begin{figure}
	\scalebox{0.25}{\includegraphics{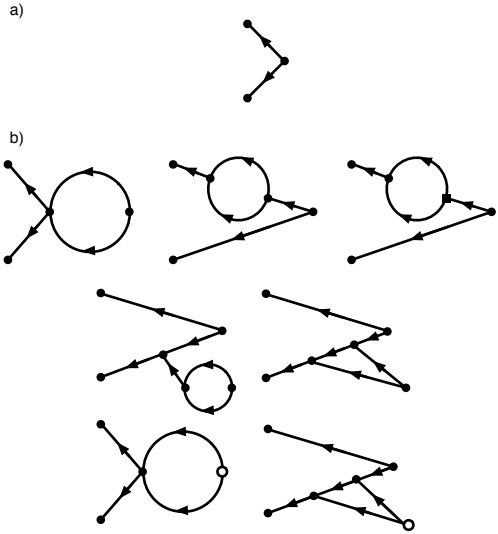}}
	\caption{Diagrams for the connected two-point function at tree level a) and to one loop b).}
	\label{fig:connectedCorrelation}
\end{figure}

The connected correlation function (cumulant function) is given by
\begin{equation}
	C(x_1,t_1 ;x_2, t_2) = \langle \psi(x_1, t_1) \psi(x_2,t_2) \rangle \label{eq:connected2}
\end{equation}
 This
is equivalent to (\ref{eq:connected}), which was computed in Ref.~\cite{hildebrand} when $t_1 = t_2$.  If the initial phases are uncorrelated (i.e. $C(x_1,0;x_2,0) = 0$) then at tree level $C(x_1,t_1;x_2,t_2)$ is given by the diagram shown in Figure~\ref{fig:connectedCorrelation} a).  It is comprised of vertex III (see Figure~\ref{fig:feynman}) combined with a bare propagator on each arm.
For general (i.e. not odd)  $f(\theta)$, the diagram in Figure~\ref{fig:connectedCorrelation} a) actually corresponds to two different terms because the arms of vertex III can be interchanged, giving two terms in equation~(\ref{eq:C_soln}) below.  Unlike, conventional field theory, these interchanges are not symmetric.    These two terms are equal when $f(\theta)$ is odd.  More generally other vertices do not exhibit the symmetries typical of Feynman diagrams even for odd $f(\theta)$. 
 Applying the Feynman rules then gives at tree level
\begin{eqnarray}
\lefteqn{C(x_1,t_1;x_2,t_2)} \nonumber \\
&=&-\frac{K}{N}\frac{1}{(2\pi)^2} \int_{-\infty}^{\infty}d\omega_1'd\omega_2'\int_{0}^{2\pi}d\theta_1'd\theta_2'\int_{t_0}^t dt'  \nonumber \\
&\times& P_0(x_1,x_1',t_1-t')P_0(x_2,x_2',t_2-t')\nonumber \\
&\times&  [\frac{\partial}{\partial \theta'_1}f(\theta'_2-\theta'_1)
  +\frac{\partial}{\partial
    \theta'_2}f(\theta'_1-\theta'_2)]g(\omega'_1)g(\omega'_2) \label{eq:C_soln}  
\end{eqnarray}
where $t = \min(t_1,t_2)$.
This is essentially identical to the  ansatz used in  Ref.~\cite{hildebrand} for the solution of the second moment equation in the BBGKY hierarchy (\ref{eq:C_rho1}).  
For $f(\theta) = \sin \theta$ and $t_1 > t_2$,  Fourier transforming Eq.~(\ref{eq:C_soln}) and inserting the bare propagator  from Eq.~(\ref{eq:proptree}) (after an inverse Laplace transformation) gives
\begin{widetext}
\begin{eqnarray}
	C_{1}(\omega_1, t_1, \omega_2, t_2)&=&  \frac{K}{N}\frac{1}{4\pi^2}g(\omega_1)g(\omega_2)   \left [  \frac{(i\omega_1 + \frac{K_c}{2})(-i\omega_2 + \frac{K_c}{2})}{(i\omega_1 + \frac{K_c}{2} - \frac{K}{2})(-i\omega_2 + \frac{K_c}{2} - \frac{K}{2})} \right.  \nonumber \\
	&&\times \left (\frac{e^{(i(\omega_1  - \omega_2) t_2)}}{i(\omega_1 - \omega_2)} - \frac{1}{i(\omega_1 - \omega_2)} \right) e^{i\omega_1(t_1 - t_2)} \nonumber \\
	&+& \frac{K}{2} \frac{ (  i \omega_2 - \frac{K_c}{2})}{(\frac{K}{2} - \frac{K_c}{2} - i \omega_1)((\frac{K_c}{2} - \frac{K}{2})^2 + (\omega_2)^2)} \left ( e^{-(\frac{K_c}{2} - \frac{K}{2} +i \omega_2)(t_2)} - 1\right) e^{-(\frac{K_c}{2} - \frac{K}{2})(t_1 -t_2)}\nonumber \\
	&+& \frac{K}{2} \frac{ (  -i \omega_1 - \frac{K_c}{2})}{(\frac{K}{2} - \frac{K_c}{2} + i \omega_2)((\frac{K_c}{2} - \frac{K}{2})^2 + (\omega_1)^2)} \left ( e^{-(\frac{K_c}{2} - \frac{K}{2} -i \omega_1)t_2} - 1\right) e^{-i\omega_1 (t_1 - t_2)} \nonumber \\
	&+&\left .  \frac{K^2}{4(K-2\frac{K_c}{2})} \frac{1}{(\frac{K}{2} - \frac{K_c}{2} - i \omega_1)(\frac{K}{2} - \frac{K_c}{2} + i \omega_2)} \left ( e^{-(K_c - K)t_2} -1 \right)\right ] e^{-(K_c - K))(t_1 - t_2)}\label{eq:cor} \nonumber \\
\end{eqnarray}
The equal time correlator is given by ($t_1 = t_2 = \tau$):
\begin{eqnarray}
	C_{1}(\omega_1, \omega_2, \tau)&=&  \frac{K}{N}\frac{1}{4\pi^2}g(\omega_1)g(\omega_2)   \left [  \frac{(i\omega_1 + \frac{K_c}{2})(-i\omega_2 + \frac{K_c}{2})}{(i\omega_1 + \frac{K_c}{2} - \frac{K}{2})(-i\omega_2 + \frac{K_c}{2} - \frac{K}{2})} \left (\frac{e^{(i(\omega_1  - \omega_2) \tau)}}{i(\omega_1 - \omega_2)} - \frac{1}{i(\omega_1 - \omega_2)} \right) \right . \nonumber \\
	&+& \frac{K}{2} \frac{ (  i \omega_2 - \frac{K_c}{2})}{(\frac{K}{2} - \frac{K_c}{2} - i \omega_1)((\frac{K_c}{2} - \frac{K}{2})^2 + (\omega_2)^2)} \left ( e^{-(\frac{K_c}{2} - \frac{K}{2} +i \omega_2)\tau} - 1\right) \nonumber \\
	&+& \frac{K}{2} \frac{ (  -i \omega_1- \frac{K_c}{2})}{(\frac{K}{2} - \frac{K_c}{2} + i \omega_2)((\frac{K_c}{2} - \frac{K}{2})^2 + (\omega_1)^2)} \left ( e^{-(\frac{K_c}{2} - \frac{K}{2} -i \omega_1)\tau} - 1\right) \nonumber \\
	&+&\left .  \frac{K^2}{4(K-2\frac{K_c}{2})} \frac{1}{(\frac{K}{2} - \frac{K_c}{2} - i \omega_1)(\frac{K}{2} - \frac{K_c}{2} + i \omega_2)} \left ( e^{-(K_c - K)\tau} -1 \right)\right ] \label{eq:corEqualTime}
\end{eqnarray}
\end{widetext}
$C_{-1} = C_1^\star$ and the other modes vanish.  Note that the initial condition $C_1(\omega, \omega', 0) = 0$ is satisfied and that the time constants and frequencies which appear are every possible way of pairing those from the tree level propagator.

For illustrative purposes, Figure~\ref{fig:connectedCorrelation} b) shows the one loop diagrams which contribute to $C$; these diagrams are $O(1/N^2)$.  We should note here the special role played by the diagrams with initial state terms, in particular the vertex proportional to $\tilde{\psi}(\theta, \omega, t_0)^2$.  This diagram evaluates to  exactly the same result as (\ref{eq:cor}), with an additional factor of $-1/N$.  It serves to provide the proper normalization for the two point function, which should go as $(N-1)/N$ since the self-interaction (diagonal) terms are not included.  The other diagrams diverge faster as one approaches criticality ($K = K_c$). They are of negligible importance at small coupling but become increasingly important near the onset of synchrony.

\subsection{Order Parameter Fluctuations}
We now compute the fluctuations in the order parameter $Z$ given in Eq.~(\ref{eq:orderp}).  The variance of $Z$  (second moment $\langle Z \bar{Z} \rangle$) is given by
\begin{eqnarray}
\langle Z \bar{Z} \rangle &=& \langle r^2 (t)\rangle \nonumber \\
&=&  \int d\omega d\omega' d\theta d\theta' \langle \eta(\omega,\theta, t) \eta( \omega', \theta',t)\rangle e^{i(\theta - \theta')}  \nonumber \\
\label{eq:flucs}
\end{eqnarray}
 Using equation~(\ref{eq:2ndmoment})  in equation~(\ref{eq:flucs}) gives
\begin{equation}
\langle r^2(t) \rangle = \int d\omega d\omega' d\theta d\theta' C(x,x',t) e^{i(\theta - \theta')} +\frac{1}{N}
\label{eq:r2gen}
\end{equation}
since in the incoherent state $\rho(x,t)=g(\omega)/2\pi$ is independent of $\theta$, so that $\langle Z \rangle =0$.
Hence
\begin{equation}
\langle r^2(\tau) \rangle = 4\pi^2 \int d\omega d\omega' C_{-1}(\omega, \omega', \tau) + \frac{1}{N}
\label{eq:r2genF}
\end{equation}
which evaluates to \cite{hildebrand}
\begin{eqnarray}
	\langle r^2(\tau) \rangle &=& \frac{2}{iKN\pi} \int_{{\cal C}} ds \frac{\Lambda_1(s - s_0) - 1}{\Lambda_1(s-s_0)} \nonumber \\
	&\times& {\rm Res}\left [ \frac{-1}{\Lambda_1(s)}\right ]_{s = s_0} \frac{1}{s}e^{s\tau} +\frac{1}{N}
	\label{eq:r2spec}
\end{eqnarray}
The time evolution of $\langle r^2 \rangle$ is then determined by the poles of $\Lambda_1(s)$.  
As an example, consider  $f(\theta) = \sin \theta$ and $g(\omega)$ a Lorentz distribution (see Appendix~\ref{app:lorentz}); we have the result
\begin{equation}
	\langle r^2(\tau) \rangle = \frac{1}{N} \frac{K_c}{K_c - K} - \frac{1}{N} \frac{K}{K_c - K} e^{-( K_c - K)\tau} \label{eq:r2}
\end{equation}
where $K_c = 2\gamma$.  Note that this diverges as $\tau \rightarrow \infty$ for $K=K_c$.  In the mean field limit $N\rightarrow\infty$, $\langle r^2\rangle=0$ as expected.
 As was shown in Ref~\cite{hildebrand}, the tree level calculation adequately captures the fluctuations except near  the  onset of synchrony ($K=K_c$). 
An advantage of the field theoretic formalism is that it allows us to approach even higher moments without needing to worry about the moment hierarchy.  In particular, for $f(\theta) = \sin \theta$ it is  straightforward to show that higher cumulants, such as $\langle (Z\bar{Z})^2 \rangle - \langle Z\bar{Z} \rangle^2$, must be zero at tree level because of rotational invariance (more precisely, any cumulant of $Z$ higher than quadratic).  The cumulants are given by graphs which are connected.  Vertex III produces two lines with wave numbers $n = \pm 1$.  Additionally, the IV and V vertices impose a shift in wave number, whereas $Z$ and $\bar{Z}$ project onto $\pm 1$.  In order to calculate these higher fluctuations it is necessary to go to the one loop level.  Note that this does \emph{not} imply that the higher cumulants of $\eta$ are zero.  Figure \ref{fig:connectedCorrelation} b) gives the diagrams for the correlation function at one loop.  The one loop calculation would also give a better estimate for $\langle r^2\rangle$,   especially nearer to criticality.  The non-interacting distribution is Gaussian, with non-Gaussian behavior growing as one approaches criticality.


\section{Linear Stability and Marginal Modes}
\label{sec:marginal}
We analyze linear stability by convolving the linear response or propagator with an initial perturbation. Although we are free in our formalism to consider arbitrary perturbations, we will consider two specific  kinds for illustrative purposes.  In the first case we perturb only the angular distribution.  In the second case we consider a perturbation which fixes one oscillator to be at a given angle $\theta$ and frequency $\omega$ at a given time $t$.  We first calculate the results at tree level which reproduces  the mean field theory results  of Ref.~\cite{strogMiro}.  In particular, we arrive at the same spectrum and Landau damping results of Refs.~\cite{strogMiro} and \cite{strogMiroMatt}.  We then define and calculate the operator $\Gamma$ (an extension of $\Gamma_0$) to  one loop order which ultimately allows us to calculate the corrections to the spectrum to order $1/N$.

\subsection{Mean field theory}

The bare propagator which is the full linear response for mean field theory is given by Eq.~(\ref{eq:proptree}).  The zeroes of the operator $\Gamma$ with respect to $s$  specify the spectrum of the linear response.  There is a set of marginal modes (continuous spectrum) along the imaginary axis spanning an interval given by the support of $g(\omega)$.  There is also a set of discrete modes given by the zeros of the dielectric function $\Lambda_n(s)$ as was found in Ref.~\cite{strogMiro}, aside from the issue of the analytic continuation of $\Lambda_n(s)$.   

However, even though there are marginally stable modes, the order parameter $Z$ can still decay to zero due to a generalized Landau damping effect as was shown in Ref.~\cite{strogMiroMatt}. 
Consider a generalization of the order parameter
\begin{equation}
	Z_n(t) = \frac{1}{N}\sum_j e^{in\theta_j},
\end{equation}
which represents Fourier modes of the density integrated over all frequencies:
\begin{eqnarray}
	Z_n(t) &=& \int d\theta d\omega \eta(\theta, \omega, t) e^{in\theta} 
	\end{eqnarray}
	and hence
	\begin{eqnarray}
	\langle Z_n(t) \rangle &=& \int d\theta d\omega \rho(\theta, \omega, t) e^{in\theta} \nonumber \\ &=& 2\pi \int d\omega \rho(-n,\omega,t).
\end{eqnarray}
In the incoherent state, $\rho(\theta,\omega,t)$ is independent of $\theta$ so $\langle Z_n \rangle$ is zero for $n>0$.  The density response $\delta\rho(\theta,\omega,t)$ to an initial perturbation $\delta \rho(\theta,\omega,0)$ is given by
 \begin{equation}
	\delta \rho(\theta,\omega, t) = N\int_{-\infty}^{\infty}\int_0^{2\pi} P_0(x,x', t)  \delta\rho(\theta',\omega',0) d\theta' d\omega' 
\label{eq:rhoprop}
\end{equation}
Recall from the definition of the action (\ref{eq:action}),  the propagator operates on an initial condition defined by $N\rho_0$.
The perturbed order parameter thus obeys
\begin{equation}
	\delta \langle Z_n(t) \rangle = \int d\theta d\omega \delta\rho(\theta, \omega, t) e^{in\theta}.
\label{eq:genop}
\end{equation}
We will show that for any initial condition involving a smooth distribution in frequency and angle, $\delta \langle Z_n(t) \rangle$ will decay to zero.  However, for non-smooth initial perturbations, $\delta\langle Z_n(t) \rangle$ will not decay to zero but will oscillate.

We first consider an initial   perturbation of  the form
\begin{equation}
	\delta \rho(\theta, \omega, 0) =   g(\omega) c(\theta)
	\label{eq:perturb1}
\end{equation}
where
\begin{equation}
	\int_0^{2 \pi} c(\theta) d\theta = 0
\end{equation}
Inserting into (\ref{eq:rhoprop}) yields
\begin{equation}
	\delta \rho(\theta,\omega, t) = N\int_{-\infty}^{\infty}\int_0^{2\pi}P_0(x,x', t)   c(\theta') g(\omega') d\theta' d\omega' 
\label{eq:delrhot}
\end{equation}
which  is consistent with the perturbation considered in Ref.~\cite{strogMiro}.
Taking the Laplace transform of (\ref{eq:delrhot}) gives
\begin{eqnarray}
	\delta \tilde{\rho}_n (\omega, s) = N c_n \int_{-\infty}^{\infty}\tilde{P}_0(n,\omega, \omega', s)   g(\omega') d\omega' 
\end{eqnarray}
Using the tree level propagator (\ref{eq:proptree}), we can show that
\begin{eqnarray}
	\int_{-\infty}^{\infty} \tilde{P}_0(n,\omega, \omega', s) g(\omega')d\omega' = \frac{1}{2\pi N} \frac{g(\omega)}{s + in\omega} \frac{1}{\Lambda_n(s)}
\end{eqnarray}
where $\Lambda_n(s)$ is given in (\ref{eq:lambda}).  Hence
\begin{equation}
	\delta \tilde{\rho}(n,\omega, s) =\frac{c_n}{2\pi } \frac{g(\omega)}{s + in\omega} \frac{1}{\Lambda_n(s)}
	\label{eq:angPertLap}
\end{equation}
From Eq.~(\ref{eq:angPertLap}), we see that the continuous spectrum is given by $in\omega$ and the discrete spectrum by the zeros of $\Lambda_n(s)$.
If we define $b_n(\omega,t) = \delta\rho(n,\omega, t)/(Nc_ng(\omega))$ then  using (\ref{eq:delrhot}) and (\ref{eq:proptree}) we can show 
\begin{eqnarray}
	&&\left [ \frac{\partial}{\partial t}  + i n \omega \right ] {b}_n(\omega,  t) \nonumber \\
	&+& in K  f(-n)   \int_{-\infty}^{\infty}b_n(\omega_1,  t) g(\omega_1) d\omega_1 \nonumber \\
	&=&\frac{1}{2\pi N} \delta(t) 
\label{eq:smlin}
\end{eqnarray}
which is equivalent to the linearized perturbation equation derived by Strogatz and Mirollo \cite{strogMiro}, with the exception that (\ref{eq:smlin}) includes the effects of the initial configuration through the source term proportional to $\delta(t)$.

Inserting 
(\ref{eq:angPertLap}) into the the Laplace transform of (\ref{eq:genop}) yields
\begin{eqnarray}
	\delta \langle \tilde{Z}_n(s) \rangle &=&c_{-n} \int \frac{g(\omega)}{s - in\omega}d\omega \frac{1}{\Lambda_{-n}(s)} \nonumber \\
	&=& c_{-n} \frac{\Lambda_{-n}(s) - 1}{-i n K f(n)}\frac{1}{\Lambda_{-n}(s)} .
\end{eqnarray}
For $f(\theta) = \sin \theta$,  $f(\pm 1) = \mp i/2$ which leads to
\begin{equation}
	\delta \langle \tilde{Z}_1(s) \rangle =c_{-1} \frac{\Lambda_{-1}(s) - 1}{ K/2}\frac{1}{\Lambda_{-1}(s)} .
\end{equation}
We note that $\delta Z_1(s)$ is identical to what was calculated in Ref.~\cite{strogMiroMatt} in which it was shown that  $\delta Z_1(t)\rightarrow 0$  as $t\rightarrow\infty$.  Even in the presence of marginal modes, the order parameter decays to zero through dephasing of the oscillators.  This dephasing effect is similar to Landau damping in plasma physics.

We can see this explicitly for the case of the Lorentz distribution
\begin{equation}
	g(\omega) = \frac{1}{\pi} \frac{\gamma}{\gamma^2 + \omega^2}
\end{equation}
From (\ref{eq:lambda}) we can calculate
\begin{equation}
	\Lambda_{\pm 1}(s) = \frac{s + \gamma - \frac{K}{2}}{s + \gamma}
\end{equation}
and $\Lambda_n(s) = 1$ for $n\ne \pm 1$.  The zero of $\Lambda_{\pm 1}(s)$  is at $s_{\pm 1} = - (\gamma - K/2)$, which provides a critical coupling 
\begin{equation}
	K_c = 2\gamma
\end{equation}
above which the system begins to synchronize.  The incoherent state is reached when $K < K_c$, which gives $s_{\pm 1}< 0$.  
Thus
\begin{eqnarray}
	\delta \langle Z_{\pm 1} \rangle &=& c_{\mp 1} e^{-\left ( \gamma - \frac{K}{2} \right) \tau} \nonumber \\
	\delta \langle Z_{n \ne \pm 1} \rangle &=& c_{-n} e^{-|n|\gamma\tau}
\end{eqnarray}
Hence angular perturbations decay away in the order parameter.  


Landau damping due to dephasing is sufficient to describe the relaxation of $Z_n(t)$ to zero for a smooth perturbation.  However, for non-smooth perturbations, this may not be true.  
Consider the linear response to a stimulus consisting of perturbing a single oscillator to have initial position $\theta_0$ and frequency $\omega_0$:
\begin{equation}
	\rho_0(\theta, \omega)  = \frac{1}{N} \left [ (N-1) \frac{g(\omega)}{2\pi} + \delta(\theta - \theta_0) \delta(\omega - \omega_0) \right ]
\end{equation}
and so the initial perturbation is
\begin{equation}
	\delta \rho_0(\theta, \omega)  = \frac{1}{N} \left [ - \frac{g(\omega)}{2\pi} + \delta(\theta - \theta_0) \delta(\omega - \omega_0) \right ]
\end{equation}
Inserting into (\ref{eq:rhoprop}) gives the time evolution of this initial perturbation 
\begin{equation}
	\delta \rho(\theta, \omega, t) = -\frac{1}{N}\frac{g(\omega)}{2\pi} + P(\theta, \omega, t; \theta_0, \omega_0, t')
	\label{eq:singleOPert}
\end{equation}
 Substituting into (\ref{eq:genop}) and taking the Lapace transform gives
\begin{eqnarray}
	\delta \langle  \tilde{Z}_n(s) \rangle &=& 2\pi \int d\omega \delta \tilde{\rho}_{-n}(\omega, s) \nonumber \\
	&=& 2\pi\int d\omega \tilde{P}(-n, \omega, \omega_0, s) \nonumber \\
	&=&  \frac{1}{N} \frac{1}{s - in\omega_0} \frac{1}{\Lambda_{-n}(s)}
	\label{eq:singleOPertZ}
\end{eqnarray}   
There are therefore two modes in $\delta Z_n(t)$, one which decays due to dephasing (determined by the zero of $\Lambda_{-n}(s)$) and one which oscillates at frequency $\omega_0$.  Thus, for this perturbation, the tree level prediction is that $\delta\langle Z \rangle$ is not zero but oscillates.
Inverse Laplace transforming~(\ref{eq:singleOPertZ}),  gives the time dependence of the order parameter
\begin{eqnarray}
	\delta \langle Z_1(t) \rangle &=& \frac{e^{i\theta_0}}{ N} \frac{1}{{\omega_0^2 + \left (\gamma - \frac{K}{2}\right)^2} }\nonumber \\
	&\times& \left [ \left(\gamma\left(\gamma - \frac{K}{2}\right) + \omega_0^2 - \frac{K}{2}i \omega_0\right )e^{-i\omega_0 t}\right. \nonumber \\
	&-&\left.\left(-i\omega_0 + \gamma - \frac{K}{2}\right) \frac{K}{2} e^{-(\gamma - \frac{K}{2}) t} \right ]
\label{eq:z1t}
\end{eqnarray}
The perturbed oscillator  has phase $\theta_0 - \omega_0 t$.  It can always be located; no information is lost as the time evolution progresses.  Hence,  for a single oscillator perturbation, the tree level calculation predicts that the order parameter will not decay to zero.  In the next section, we show that to the next order in the loop expansion, which accounts for finite size effects,  the marginal modes are moved off of the imaginary axis stabilizing the incoherent state, and the order parameter for a single oscillator perturbation decays to zero.

\subsection{Finite size effects}
Let us  define a generalization of the operator $\Gamma_0$.
	\begin{equation}
		\Gamma \cdot P = \frac{1}{N} \delta(x - x') \delta(t - t')
	\end{equation}
	where $P$ without a subscript denotes the full propagator and the operator $\Gamma$ is the functional inverse of $P$. 
We can estimate the effect of  finite size on the stability of the incoherent state by calculating the one loop correction to the operator $\Gamma$.  We will see that the one loop correction produces the effect, among others, of adding a diffusion operator to $\Gamma$, which is enough to stabilize the continuum of marginal modes because the continuous spectrum is pushed off the imaginary axis by an amount proportional to the diffusion coefficient.  

\begin{figure}
	\scalebox{0.25}{\includegraphics{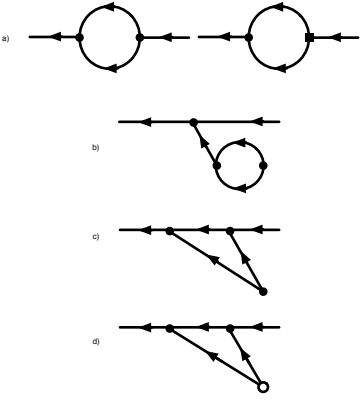}}
	\caption{The diagrams contributing to the propagator at one loop order, organized by topology.  We consider d) to be of different topology than c) because it is equivalent to a tree level diagram with an additional factor of $1/N$, due to the initial state vertex.}
	\label{fig:oneloopprop}
\end{figure}

\begin{figure}
	\scalebox{0.25}{\includegraphics{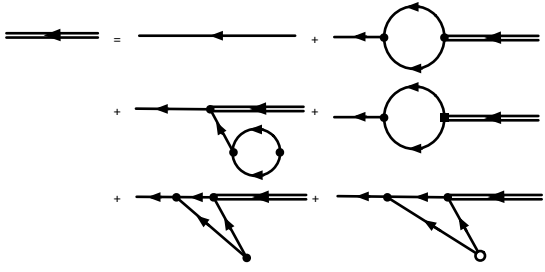}}
	\caption{Diagrammatic equation for the propagator.  The double lines represent the summation of the entire series in $1/N$ for the propagator.}
	\label{fig:fullProp}
\end{figure}

We calculate the correction to $\Gamma$  to one loop order.  The propagator is represented by diagrams with one incoming line and one outgoing line.  There are four groups of diagrams which contribute to the propagator at one loop order.  They are shown in Figure~\ref{fig:oneloopprop} and are labeled by a), b), c), and d).
Using these graphs to calculate the propagator to order $1/N$ is not sufficient to demonstrate the behavior of the spectrum to order $1/N$.  
However, we can use these graphs to construct an approximation of $\Gamma$ to order $1/N$ and derive the spectrum from this.  
If we denote the full propagator (i.e. the entire series in $1/N$ for $P$) by a double line, we can approximate the  full propagator recursively by the diagrammatic equation shown in Figure~\ref{fig:fullProp}.  The only terms which are neglected in this relation are those which are from two loop and higher graphs and therefore would contribute $O(1/N^2)$ to $\Gamma$.  Readers familiar with field theory will note that we are simply calculating  the two point proper vertex, which is the inverse of the full propagator, to one loop order.  

If we act on both sides of this equation with $\Gamma_0$ (the operator whose inverse is the tree level propagator), we arrive at an equation of the form:
\begin{equation}
	\Gamma_0 \cdot P = \frac{1}{N} \delta(x - x') \delta(t - t') - \Gamma_1 \cdot P
	\label{eq:fullPropEQ}
\end{equation}
where we have implicitly defined the one loop correction to $\Gamma$, which we label $\Gamma_1$.
The action of $\Gamma_0$ converts the leftmost propagator in each diagram into a delta function, so that the delta function term in (\ref{eq:fullPropEQ}) arises from the tree level propagator line and $\Gamma_1$ is then comprised of the loop portions of the remaining diagrams (the ``amputated" graphs).
We denote the contribution to $\Gamma_1$ from each group of one loop diagrams  by $\Gamma_{1r}(\theta, \omega; \phi, \eta; t - t')$, where $r$ represents $a,b$, $c$, or $d$ indicating the group of  diagrams in question.   $\Gamma_{1b} = 0$   because the derivative coupling acts on the incoherent state, which is homogeneous in $\theta$.  The equation of motion for the one loop propagator $P_1(x,x',t)$ then has the form
\begin{equation}
\Gamma_0\cdot P_1 +  \Gamma_1\cdot P_1  = \frac{1}{N}\delta(\theta - \theta') \delta(\omega - \omega') \delta(t - t') 
\end{equation}
where $\Gamma_0\cdot P_1$ is given by
\begin{widetext}
\begin{eqnarray}
	\Gamma_0\cdot P_1
	&=& \left [ \frac{\partial}{\partial t} + \omega \frac{\partial}{\partial \theta} + K\frac{\partial}{\partial \theta}\int_{-\infty}^{\infty}\int_{0}^{2\pi}f(\theta_1-\theta)\rho(x_1,t)d\theta_1 d\omega_1\right ] P_1(x,x',t-t') \nonumber \\
&+&K\frac{\partial}{\partial \theta}\int_{-\infty}^{\infty}\int_{0}^{2\pi}f(\theta_1-\theta)\rho(x,t)P_1(x_1,x',t-t')d\theta_1 d\omega_1
\end{eqnarray}
and
\begin{eqnarray}
	\Gamma_1\cdot  P_1=& &\int_0^{2\pi} d\phi \int_{-\infty}^{\infty}d\eta \int_{t'}^{t} dt''\Gamma_{1a}(\theta, \omega; \phi, \eta; t - t'') P_1(\phi, \eta,t'';\theta', \omega'; t') \nonumber \\
	&+& \int_0^{2\pi} d\phi \int_{-\infty}^{\infty}d\eta \int_{t'}^{t} dt'\Gamma_{1c}(\theta, \omega; \phi, \eta; t - t'') P_1(\phi, \eta,t'';\theta', \omega'; t')  \nonumber \\
	&+& \int_0^{2\pi} d\phi \int_{-\infty}^{\infty}d\eta \int_{t'}^{t} dt'\Gamma_{1d}(\theta, \omega; \phi, \eta; t - t'') P_1(\phi, \eta,t'';\theta', \omega'; t') 
	\label{eq:OneLoopProp}
\end{eqnarray}
\end{widetext}
is the one loop contribution.
The kernels  $\Gamma_{1a}$, $\Gamma_{1c}$, and $\Gamma_{1d}$ are explicitly computed in
 Appendix~\ref{app:oneloop}.  
 
 The expressions for the one loop contribution to $\Gamma$ are rather complicated but several key features can be extracted, namely 1) the introduction of a diffusion operator, 2) a shift in the driving frequency, and 3) the addition of higher order harmonics to the coupling function $f$.  The diffusion operator has the effect of shifting the marginal spectrum from the imaginary axis into the left hand plane.   The effect is that the finite size fluctuations to order $1/N$ stabilize the incoherent state.  
 

 We can see these effects more easily by considering the special case of
  $f(\theta) = \sin \theta$ and $g(\omega)$ being a Lorentz distribution (see  Appendix~\ref{app:lorentz}).  
The Fourier-Laplace transformed equation of motion for the one loop propagator has the form
\begin{eqnarray}
	&&\alpha_{\pm 1}(s;\omega) \tilde{P}_1(n, \omega, \omega', s)  \nonumber \\
	&-& \frac{K}{2}g(\omega) ( 1 + \beta_1(s;\omega)) \int d\nu \tilde{P}_1(n,\nu, \omega', s) \nonumber \\
	&=& \frac{1}{2\pi N} \delta(\omega - \omega'),\qquad n=\pm1,
\label{eq:first}
\end{eqnarray}
\begin{eqnarray}
	&&\alpha_{\pm 2}(s; \omega) \tilde{P}_1(n,\omega, \omega', s) \nonumber \\
	&-& \frac{K}{2}g(\omega) \int d\nu  \beta_{\pm 2}(s;\omega, \nu) \tilde{P}_1(n, \nu, \omega',s) \nonumber\\
	&=& \frac{1}{2\pi N} \delta(\omega - \omega'),\qquad n=\pm2,
\label{eq:second}
\end{eqnarray}
and
\begin{eqnarray}
	\alpha_n(s;\omega) \tilde{P}_1(n,\omega, \omega', s) 
	= \frac{1}{2\pi N} \delta(\omega - \omega'),\qquad |n|>2 \nonumber \\
	\label{eq:ng2Prop}
\end{eqnarray}
where
\begin{widetext}
\begin{eqnarray}
	\alpha_n(s;\omega)&=& \left [ s + i n\omega  + \frac{K^2}{4N} \sum_{m=\pm 1}  \frac{1}{s + \gamma - \frac{K}{2} + i(m+n)\omega}     \left ( n(2m+n) + n(m+n)\frac{K}{2\gamma - K} \right ) \right ] \nonumber \\
	\beta_{\pm 1}(s;\omega) &=&
	 -\frac{K^2}{N}    \frac{1}{s +\gamma - \frac{K}{2} \pm 2 i \omega}
	 \frac{1}{\gamma - \frac{K}{2} \pm i\omega}\left (\frac{2\gamma - \frac{K}{2}}{2 \gamma - K} \right ) +   \frac{1}{N} \frac{ \frac{K}{2}}{ s + \gamma - \frac{K}{2}} \nonumber \\
	 \beta_{\pm2}(s;\omega,\eta) &=& -\frac{K^2}{N}\left (  \frac{s \mp i \omega}{s \mp i\omega + \gamma - \frac{K}{2}}   \frac{1}{s \pm i (\eta - \omega)} \frac{\pm i\omega + \gamma}{\left (\gamma - \frac{K}{2} \right )^2 + \omega^2}   \right .  \nonumber \\
	&&  \left . \left . + \frac{s + \gamma - \frac{K}{2}}{s + 2\gamma - K}  \frac{1}{s + \gamma - \frac{K}{2}\pm\eta}\frac{1}{2\gamma - K} \frac{1}{-\gamma + \frac{K}{2} \mp i\omega} \frac{K}{2}  \right. \right . \nonumber \\
	&&   \left . \pm \frac{K}{2} g(\omega)\frac{1}{2\gamma - K} \frac{1}{s + \gamma - \frac{K}{2} \pm i\omega } \frac{1}{s + \gamma - \frac{K}{2} \pm i\eta } \frac{s + 2\gamma - \frac{K}{2}}{s + 2\gamma - K}\right ).
\end{eqnarray}
\end{widetext}
We can solve for $\tilde{P}_1$ using the same kind of self-consistency computation that we used for $\tilde{P}_0$.  This produces an analogously defined dialectric function, $\Lambda_n^1(s)$.
\begin{eqnarray}
	\Lambda^1_n(s) = 1 - \frac{K}{2}\int d\omega \frac{g(\omega)(1 + \beta_n(s;\omega))}{\alpha_n(s;\omega)}
	\label{eq:onelooplambda}
\end{eqnarray}

Stability of the incoherent state is determined by the  spectrum of the operator $\Gamma$. 
Analogous to tree level, the continuous spectrum is given by the zeros of $\alpha_n(s;\omega)$ and the discrete spectrum by the zeros of $\Lambda^1_n(s)$ given by  (\ref{eq:onelooplambda}).  However, at one loop order,  the expressions for $\tilde{P}_1$ represent solutions to a coupled system of equations.  Thus,  the poles of tree level are shifted, and  there are also new poles reflecting the interaction of the mean density with the two-point correlation function.  These poles will have residues of $O(1/N^2)$ owing to their higher order nature.  For $n=\pm1, \pm2$, we cannot solve for the poles exactly but we can approximate the shift in the tree level spectrum by evaluating the loop correction at the value of the  tree level pole, $s = \mp i n \omega$, which is equivalent to using the 
``on-shell" condition in field theory.  Since the higher order modes will decay faster than the tree level modes, this essentially amounts to ignoring short time scales and is similar to the Bogoliubov approximation \cite{ichimaru, nicholson}.  
The remaining effective equation for $\tilde{P}_1$ is now first order and, consequently, we can consider the spectrum of the implicitly defined operator analogous to (\ref{eq:linearO}).

The continuous spectrum consists of all the zeros of the function $\alpha_n$.  We expect a term of the form $in\omega + O(1/N)$ because of the tree level continuous spectrum.  This will govern the behavior at large times.  
In this case, the ``on-shell" condition is equivalent to Taylor expanding the loop correction via $s = in\omega + O(1/N)$ and keeping only terms which are $O(1/N)$.  This yields:
\begin{equation}
\alpha_n(s;\omega)=s + in( \omega + \delta \omega) + n^2 D
\end{equation}
where
\begin{equation}
	\delta\omega=  - \frac{K^2}{2N}\frac{\omega}{\left ( \gamma - \frac{K}{2} \right )^2 + \omega^2} \left [ \frac{4\gamma - K}{2\gamma - K} \right] 
\end{equation}
is a frequency shift and
\begin{equation}
	D=\frac{K^2}{2N}\frac{\gamma}{\left ( \gamma - \frac{K}{2} \right )^2 + \omega^2 } 
	\label{diffusioncoef}
\end{equation}
is a diffusion coefficient.  The frequency shift, which is negative, serves to tighten the distribution around the average frequency.  The diffusion operator serves to damp the modes which are marginal at tree level.  

The discrete spectrum arises from the zeroes of $\Lambda^1_n(s)$.  We can again approximate the shift in the tree level zero by using the on-shell condition.  This gives
\begin{eqnarray}
	\Lambda^1_n(s) &=& 1 - \frac{K}{2} \int d\omega \frac{g(\omega)}{s + in (\omega + \delta \omega) + n^2 D} \nonumber \\
	&\times& \left ( 1 -\frac{K^2}{N} \frac{2 \gamma - \frac{K}{2}}{2 \gamma - K} \frac{1}{ \left ( \gamma - \frac{K}{2} +in\omega \right )^2} \right. \nonumber \\
	 &+& \left.  \frac{1}{N} \frac{ \frac{K}{2}}{   \gamma - \frac{K}{2} -in\omega}\right )
\end{eqnarray}
We assume the shift will be small which allows us to write the zero of $\Lambda^1_n(s)$ as 
\begin{equation}
	s = -\left ( \gamma - \frac{K}{2} \right ) - \frac{\delta \Lambda^1_n(s_0)}{\left (\frac{d\Lambda^1_n(s_0)}{ds}\right )}
\end{equation}
where $s_0 =  -\left ( \gamma - \frac{K}{2} \right )$ and $\delta \Lambda^1_n(s)$ is the $O(1/N)$ correction to $\Lambda^1_n(s)$.
This results in
\begin{eqnarray}
	s_n &=& -\left ( \gamma - \frac{K}{2} \right ) \nonumber \\
	&+& \frac{1}{N} \frac{K}{2} \left [ \left ( \frac{K}{2\gamma - K}\right ) \frac{K\gamma}{\left ( \gamma - \frac{K}{2}\right)^2 - \gamma^2}  + \frac{6\gamma - K}{2\gamma - K} \right ]. \nonumber \\
\end{eqnarray}
Away from criticality ($K \ll 2\gamma$) or for large $N$, this correction is small.

We conclude this section by writing down an effective equation of motion for the density function which incorporates the effect of fluctuations.
Recall the first equation of the BBGKY hierarchy is
\begin{eqnarray}
\frac{\partial \rho}{\partial t}&+&\omega \frac{\partial \rho}{\partial \theta} \nonumber \\
&+&K\frac{\partial}{\partial \theta}\int_{-\infty}^{\infty}\int_{0}^{2\pi}f(\theta'-\theta)\rho(x,t)\rho(x',t)d\theta' d\omega' \nonumber\\
&=&-K\frac{\partial}{\partial \theta}\int_{-\infty}^{\infty}\int_{0}^{2\pi}f(\theta'-\theta)C(x;x',t)d\theta' d\omega' \label{eq:bbgky1-1}
\end{eqnarray}
This equation has a term (the ``collision" integral) involving the 2-point correlation function, $C$, on the right hand side.
The equation determining the tree level propagator (\ref{eq:TreeProp}) is the linearization of the first BBGKY equation with $C$ considered to be zero.  The one loop correction to this equation (\ref{eq:OneLoopProp}) incorporates the effect of the correlations on the linearization.  The diagrams in Figure~\ref{fig:oneloopprop} provide the linearization of the collision term, where $C$ is considered as a functional of $\rho$.
Using our one loop calculation, we can propose an effective density equation at one loop order
\begin{widetext}
\begin{eqnarray}
\frac{\partial \rho}{\partial t} + \Omega \frac{\partial \rho}{\partial
  \theta}  - D\frac{\partial^2 \rho}{\partial \theta^2}  
&+&K\frac{\partial}{\partial
  \theta}\int_{-\infty}^{\infty}\int_{0}^{2\pi}\sin(\theta'-\theta) \rho(\theta',\Omega',t) \rho(\theta,\Omega,t)d\theta'
  d\omega' \nonumber \\
  &=& -K_2(\omega) \frac{\partial}{\partial
  \theta}\int_{-\infty}^{\infty}\int_{0}^{2\pi}\sin(2\theta'-2\theta) \rho(\theta',\Omega',t) \rho(\theta,\Omega,t)d\theta'
  d\omega'
\label{eq:oneLoopCont} 
\end{eqnarray}
\end{widetext}
where  $\Omega =  \omega + \delta \omega$ and $D$ is given by Eq.~(\ref{diffusioncoef}).
The field  $ \rho(\theta, \Omega, t)$ is now defined in terms of the shifted frequency distribution
\begin{equation}
	G(\Omega) \approx g(\omega) ( 1 - \frac{d \delta \omega}{d \omega} )
\end{equation}
The new coupling constant $K_2(\omega)$ is $O(1/N)$ and is due solely to the fluctuations.  It arises from the term $\beta_{\pm 2}$ in the equation for $\tilde{P}_1$.  In fact, given the structure of the diagrams, it is clear that for  $O(1/N^n)$ there will be  a new coupling, $K_{n+1}$, which corresponds to a $\sin[(n+1)\theta]$ term.   In the language of field theory, all odd couplings are generated under renormalization.  

The generation of higher order couplings is especially interesting in light of the results of Crawford and Davies concerning the scaling of the density $\eta$ beyond the onset of synchronization, i.e. $\eta - \rho_0 \sim (K - K_c)^\beta$ \cite{crawford2, crawford3}.  Although our calculations pertain to the incoherent state, the fact that the loop corrections generate higher order couplings is a general feature of the bulk theory defined by the action of Eq.~(\ref{eq:action1}) as well.  Thus, we expect a crossover from $\beta = 1/2$ to $\beta = 1$ behavior to occur as $N$ gets smaller.  This is consistent with \cite{crawford2}, wherein a crossover manifested as the rate constant became smaller than the externally applied diffusion. In our case, the magnitude of the diffusion is governed by the distance to criticality and the number of oscillators.

Our proposed effective equation (\ref{eq:oneLoopCont}) is not self-consistent because
 we use the propagator to infer the form of the mean field equation.  Thus, we  neglect non-linear terms which may arise due to the loop corrections.  In addition, our calculation applies specifically to perturbations in the incoherent state.  There are likely other terms we are neglecting for both of these reasons.  The consistent approach would be to calculate the effective action to one loop order and derive the equation for $\rho$ from that. This would involve essentially the same calculation we have performed here, but for arbitrary $\rho(\theta, \omega, t)$ (i.e. we would need to solve (\ref{eq:TreeProp}) for the propagator in the presence of an arbitrary mean).  



\section{Numerical Simulations}
\label{sec:simulation}
	We compare our analytical results to simulations of single oscillator perturbations, since  this provides a direct measurement of the propagator per equation~(\ref{eq:singleOPert}).   We perform simulations of $N$ oscillators with $f(\theta) = \sin \theta$.  We fix 2\% of the oscillators at a specific angle ($\theta_0 = 0$) and driving frequency (unless $N=10$, in which case we fix a single oscillator; the plots with $N=10$ have been rescaled to match the other data).  The remaining oscillators are initially uniformly distributed over angle $\theta$ with driving frequencies drawn from a Lorentz distribution.  We measure the real part of $Z_1(t)$.  This measurement allows us to observe the behavior of the modes which are marginal at tree level. 
	
Equation~(\ref{eq:z1t}) gives the behavior of $\delta Z_1(t)$ with a single oscillator fixed at $\theta_0$ and $\omega_0$ at time $t=0$.  Recall that $Z_1 = 0$ in the incoherent state, so that we expect $\delta Z_1 \approx  Z_1$.  To tree level
\begin{widetext}
\begin{equation}
	 Z_1(t) = \frac{e^{i\theta_0}}{ N} \frac{1}{{\omega_0^2 + \left (\gamma - \frac{K}{2}\right)^2} }\left [ (\gamma(\gamma - \frac{K}{2}) + \omega_0^2 - \frac{K}{2}i \omega_0 )e^{-i\omega_0 t} -(-i\omega_0 + \gamma - \frac{K}{2}) \frac{K}{2} e^{-(\gamma - \frac{K}{2}) t} \right ]
\end{equation}
In other words, the initially fixed oscillator  has phase $\theta_0 - \omega_0 t$; no information is lost as the time evolution progresses.

Incorporating the one loop computation gives
\begin{equation}
	Z_1(t) = \frac{e^{i\theta_0}}{ N} \frac{1}{{\omega_0^2 + \left (\gamma - \frac{K}{2}\right)^2} }\left [ (\gamma(\gamma - \frac{K}{2}) + \omega_0^2 - \frac{K}{2}i \omega_0 )e^{i(\omega +\delta \omega) t -  Dt} -(-i\omega_0 + \gamma - \frac{K}{2}) \frac{K}{2} e^{s_1 t} \right ]
\end{equation}
where we have ignored a term of amplitude $O(1/N^2)$; we are only considering the contributions coming from the poles described in the previous section.  With the one loop corrections taken into account, we see that $ Z_1(t)$ relaxes back to zero as $t \rightarrow \infty$.  In the simulations, we compute the real part of $ Z_1(t)$ with $\theta_0 = 0$.  This gives
\begin{eqnarray}
	{\rm Re}( Z_1(t) ) &=& \frac{1}{ N} \frac{1}{{\omega_0^2 + \left (\gamma - \frac{K}{2}\right)^2} }\left [ \left ( \gamma (\gamma - \frac{K}{2} )  + \omega_0^2 \right ) \cos \left ((\omega_0 + \delta \omega)t \right ) e^{-Dt}  \right . \nonumber \\
	&&+\left . \frac{K\omega_0}{2} \sin\left ((\omega_0 + \delta \omega)t \right ) e^{-Dt} - \frac{K}{2}(\gamma - \frac{K}{2}) e^{s_1 t} \right ]
\end{eqnarray}
\end{widetext}
The special case of $\omega_0 = 0$ and $\theta_0 = 0$ gives:
\begin{equation}
	 Z_1(t) = \frac{1}{ N} \frac{1}{{\left (\gamma - \frac{K}{2}\right)} }\left [ \gamma e^{ -  Dt} - \frac{K}{2} e^{s_1 t} \right ]
	 \label{estimate}
\end{equation}
The imaginary part vanishes so that ${\rm Re}( Z_1(t))  =  Z_1(t)$.

\begin{figure}
	\scalebox{0.25}{\includegraphics{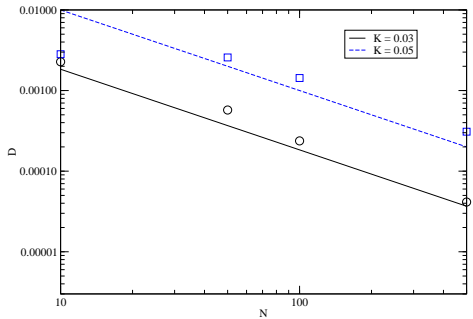}}
	\caption{The large time ($> 200 s$) decay constants with zero driving frequency for the perturbed oscillator.  Lines are the predictions given by Eq.~\ref{diffusioncoef}).  Solid line and circles represent K = 0.03. Dashed line and boxes represent K = 0.05. }
	\label{fig:decay}
\end{figure}

\begin{figure}
	\scalebox{0.25}{\includegraphics{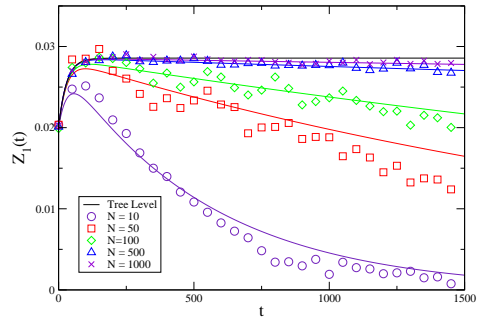}}
	\caption{$Z_1(t)$ vs. $t$ for various values of $N$ and $K= 0.3K_c$. Each graph shows  $N= \{10,50,100,500,1000\}$. Note that as $N \rightarrow \infty$ the curve approaches the tree level value.  From top to bottom:  Black line represents tree level.  X's and violet line represent N = 1000.  Triangles and blue line represent N = 500.  Diamonds and green line represent N = 100.  Boxes and red line represent N = 50.  Circles and purple line represent N = 10.}
	\label{fig:k03}
\end{figure}

\begin{figure}
	\scalebox{0.25}{\includegraphics{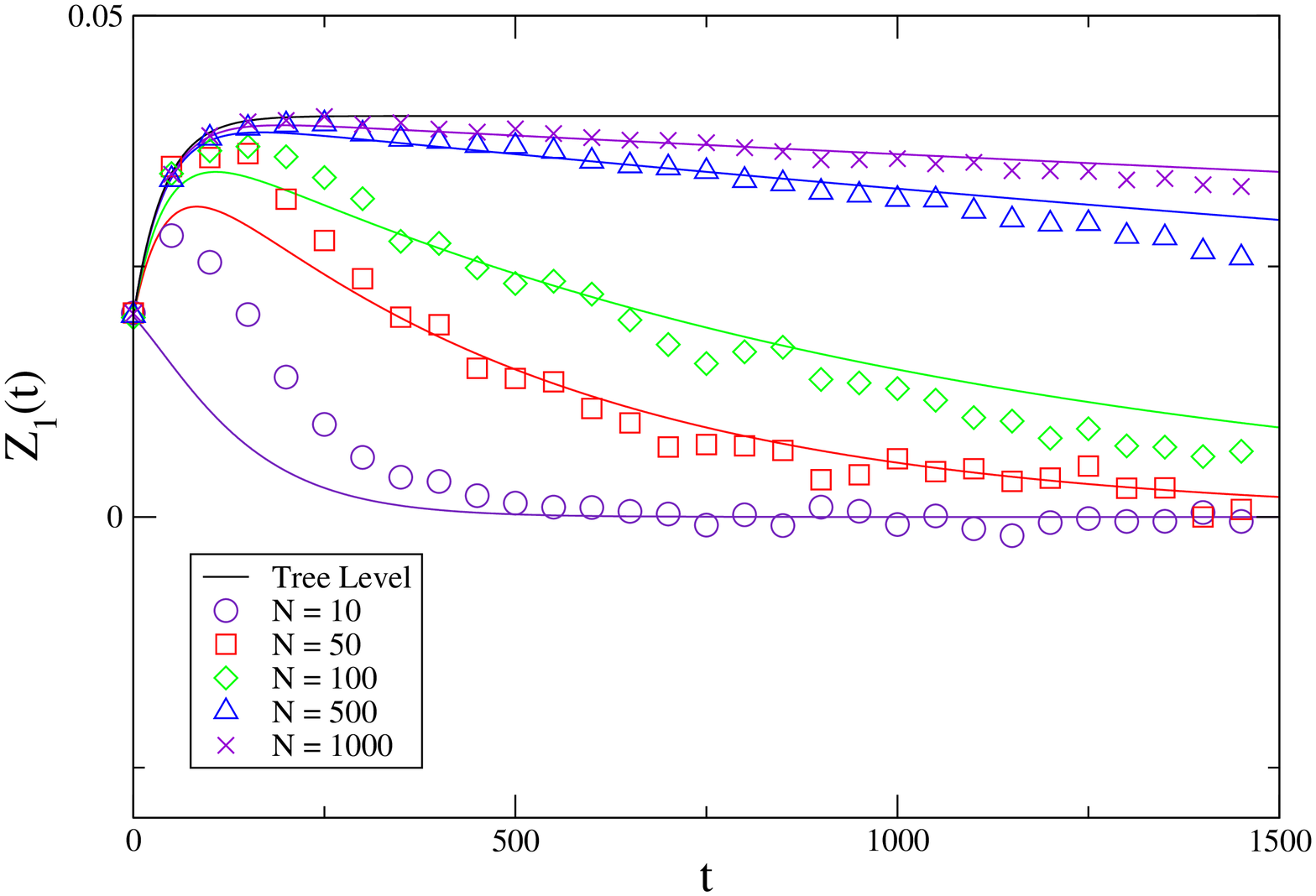}}
	\caption{$Z_1(t)$ vs. $t$ for various values of $N$ and $K= 0.5K_c$. Each graph shows  $N= \{10,50,100,500,1000\}$. Note that as $N \rightarrow \infty $ the curve approaches the tree level value. Symbols as in Figure~\ref{fig:k03}.}
	\label{fig:k05}
\end{figure}

We first compare our estimate of the diffusion coefficient $D$ given by (\ref{diffusioncoef}) with the simulations.
We plot the measured decay constant $D$  of $Z_1$  compared to the theoretical estimate of (\ref{estimate}) for the long time behavior in Figure~\ref{fig:decay}.  These data only include values of $K = 0.3K_c$ and $K=0.5K_c$ ($K_c = 2\gamma = 0.1$).  Higher values of $K$ did not yield good fits due to the neglected contributions to $Z_1$.  These decay constants are obtained via fitting the time evolution of $Z_1$ to an exponential for $t > 200 s$. 
 In both cases they behave as $1/N$ for large $N$ as predicted.  There is a consistent discrepancy likely due to rounding error after simulating for such a long period of time ($\approx 30000$ time steps).  This error appears as a small degree of noise which further damps the response, hence the decay constants appear slightly larger in Figure~\ref{fig:decay}.  This effect can be seen in Figures~\ref{fig:k03} and \ref{fig:k05} as well.  For large times, the simulation data consistently fall slightly under the analytic prediction.  Similarly, the data is noisier at large times.
 
 Figures~\ref{fig:k03} and \ref{fig:k05} show the evolution of $Z_1(t)$ over time along with the analytical predictions and the tree level result for $K = 0.3K_c$ and $K=0.5K_c$ respectively.  For $K=0.3K_c$, the prediction works quite well, with perhaps the beginning of a systematic deviation appearing at $N=10$ and $N=50$ (there is a slight initial overshoot followed by an undershoot at larger times).  This same deviation is more pronounced for $K=0.5K_c$, although the data follow the prediction quite well nonetheless.
 
\begin{figure}
		\scalebox{0.25}{\includegraphics{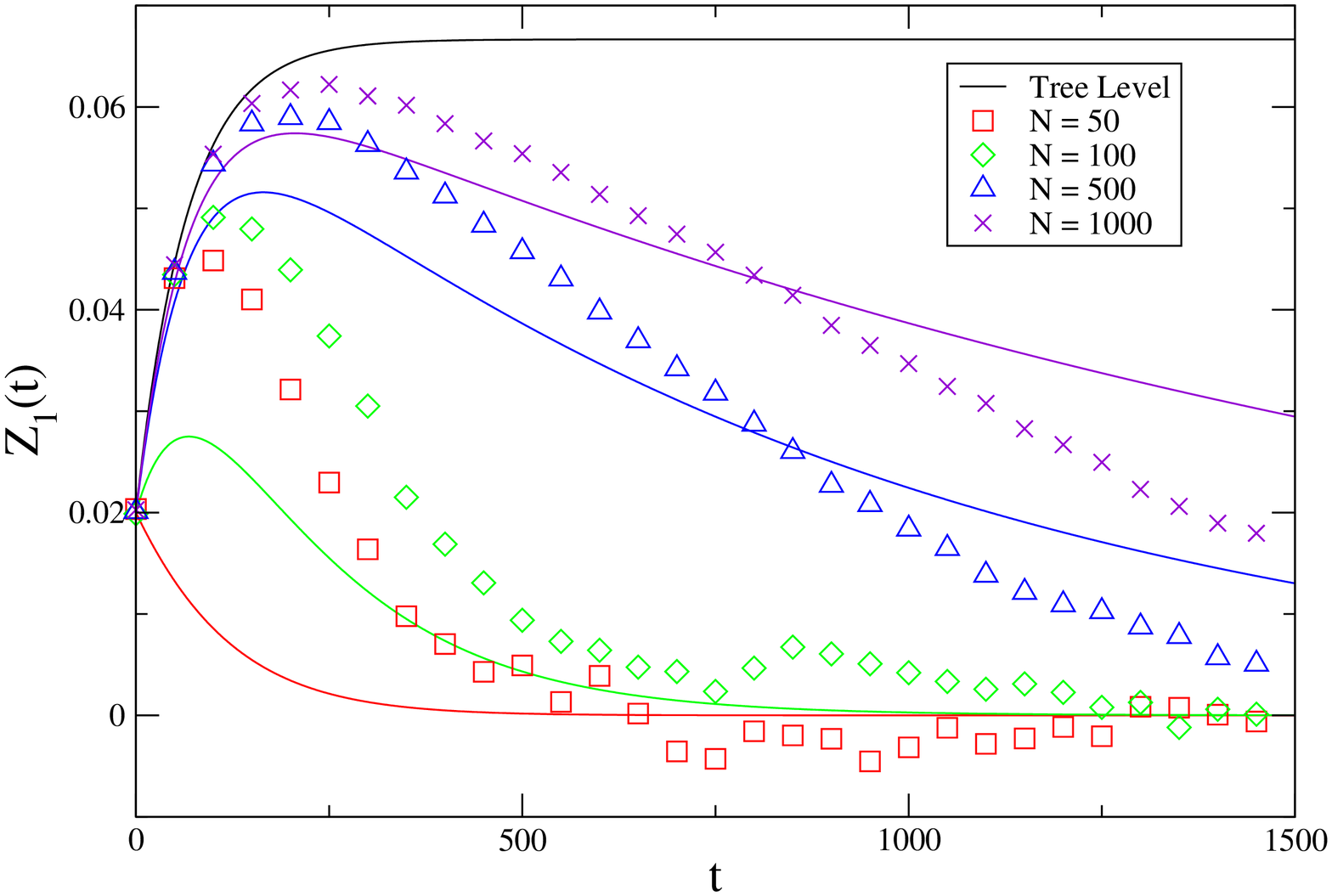}}  
		\scalebox{0.25}{\includegraphics{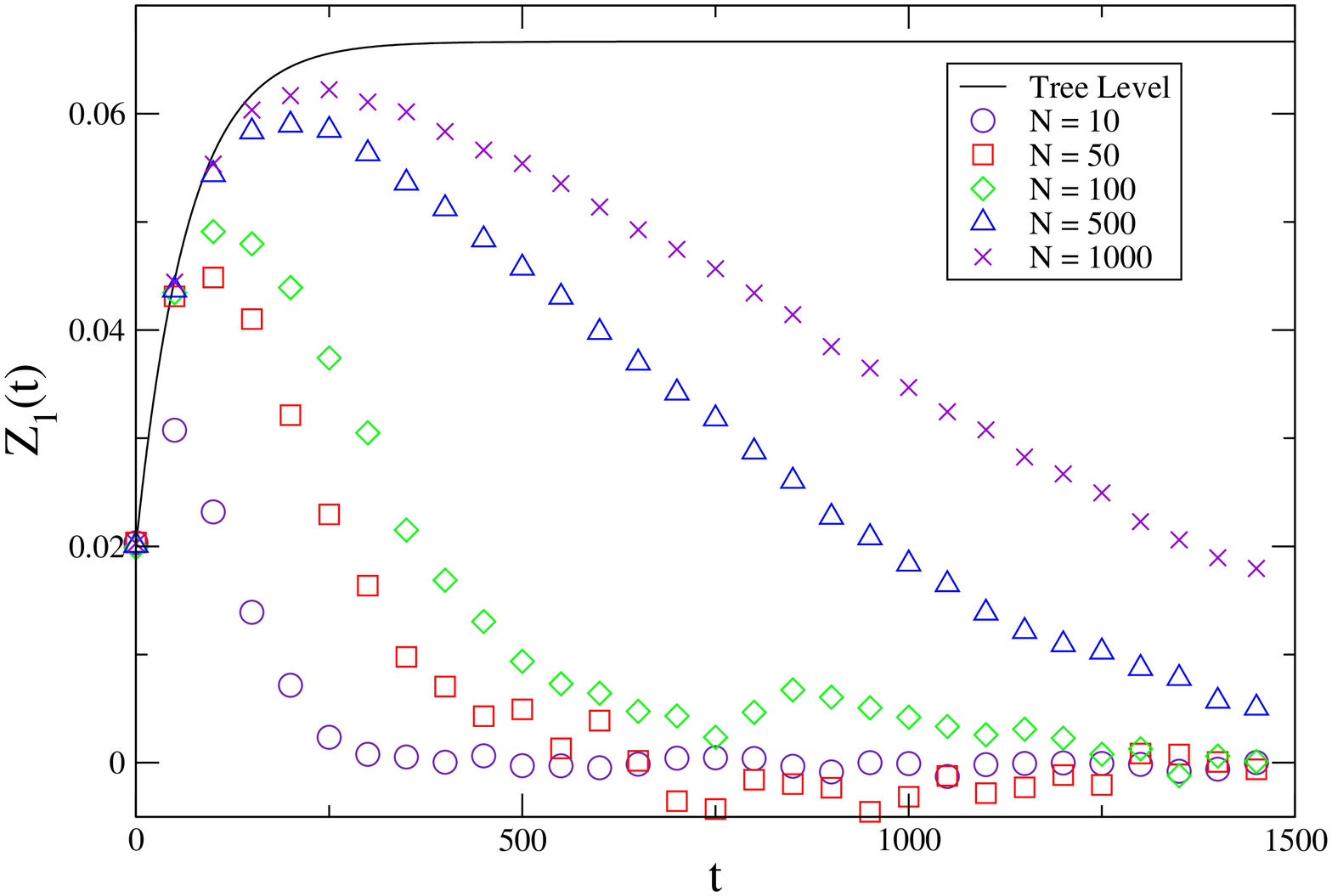}} 
	\caption{$Z_1(t)$ vs. $t$ for various values of $N$ and $K= 0.7K_c$. Each graph shows  $N= \{10,50,100,500,1000\}$. Note that as $N \rightarrow \infty $ the curve approaches the tree level value. Symbols as in Figure~\ref{fig:k03}.}
	\label{fig:k07}
\end{figure}

\begin{figure}
	\scalebox{0.25}{\includegraphics{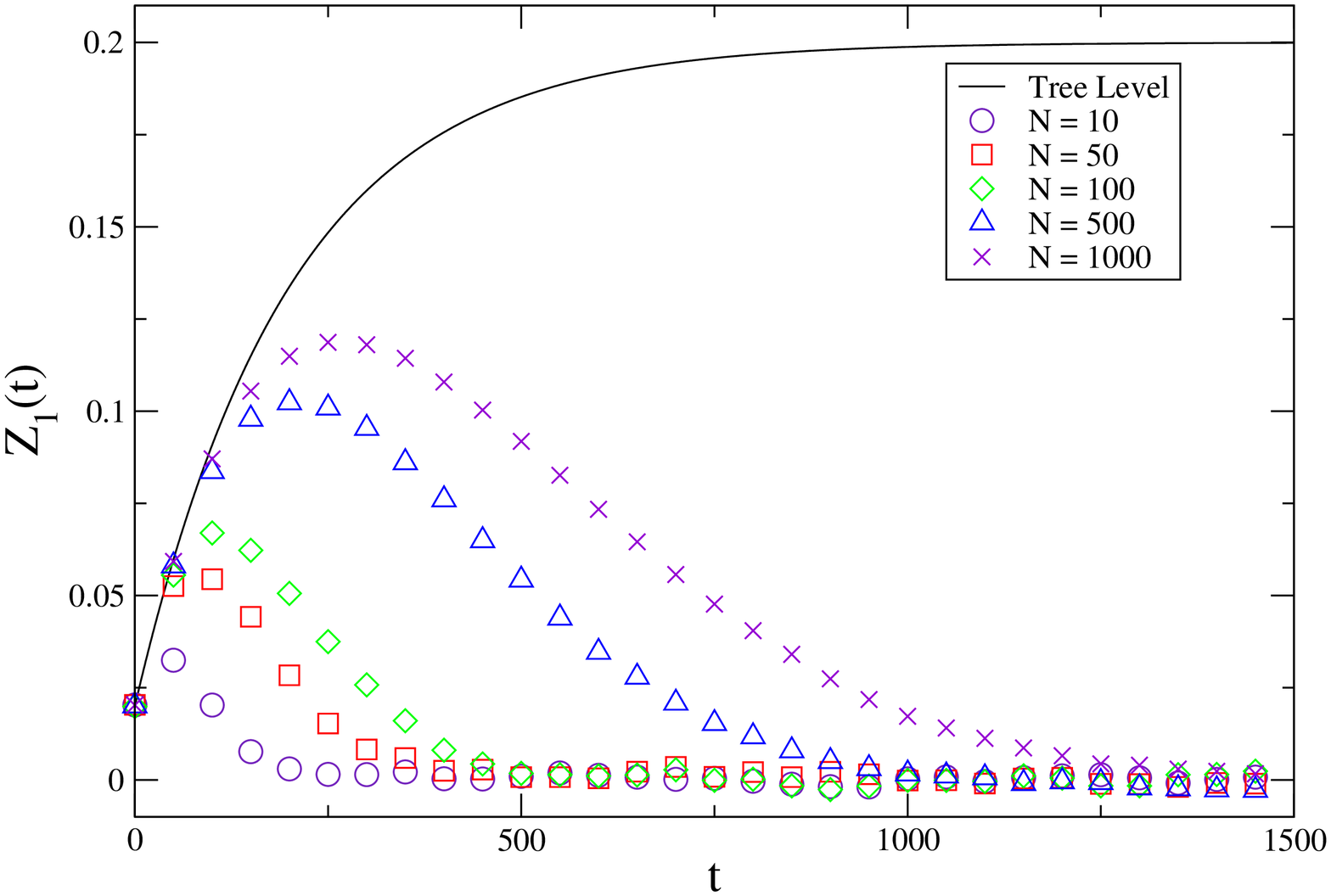}}
	\caption{$Z_1(t)$ vs. $t$ for various values of $N$ and $K= 0.9K_c$. Each graph shows  $N= \{10,50,100,500,1000\}$. Note that as $N \rightarrow \infty $ the curve approaches the tree level value. Symbols as in Figure~\ref{fig:k03}.}
	\label{fig:k09}
\end{figure}

 Consistent with our expectations from the loop expansion, as we move closer to criticality, i.e.~the onset of synchronization,  the results for $K=0.7K_c$ and $K=0.9K_c$ do not fare as well.  Figure~\ref{fig:k07} demonstrates a marked deviation from the prediction.  We have not shown analytical results for the lower values of $N$ because the deviation is so severe.  The same holds true for all the results for $K=0.9K_c$, so that we have just plotted the simulation data in Figure~\ref{fig:k09}.  The general trend of approaching the mean field result still holds.  The primary feature to take from these plots is that the fluctuations increase the decay constant.  The closer to criticality, the more important the fluctuations and the faster the decay, hence the systematic undershoot which grows as one nears criticality.  The fastest relaxation to the incoherent state appears at high $K$ for a given $N$ and at low $N$ for a given $K$, either limit results in increased effects from fluctuations.  It would be necessary to carry the loop expansion to two or more loops in order to obtain good matches with these data.

\begin{figure}
	\scalebox{0.25}{\includegraphics{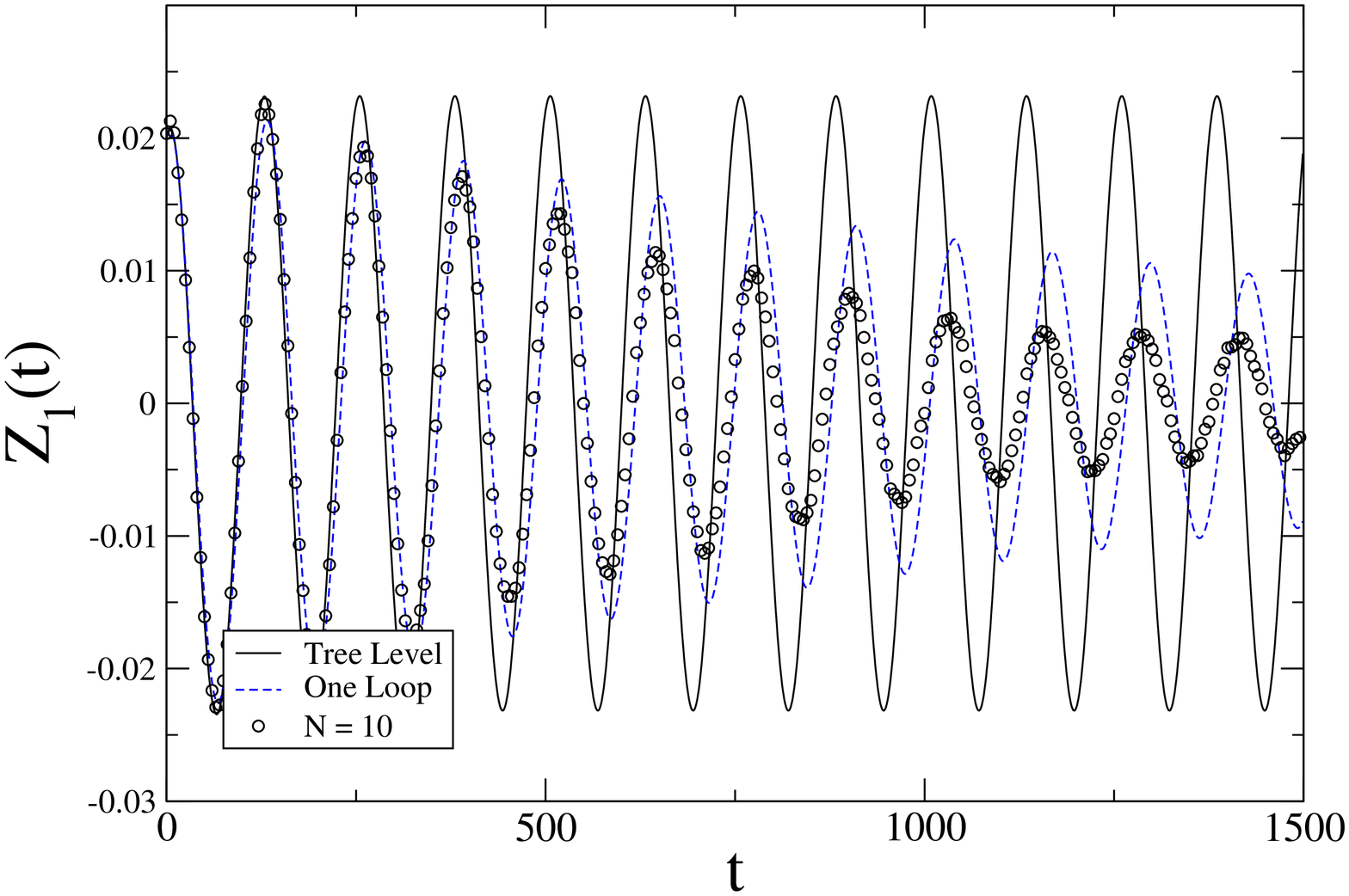}}
	\scalebox{0.25}{\includegraphics{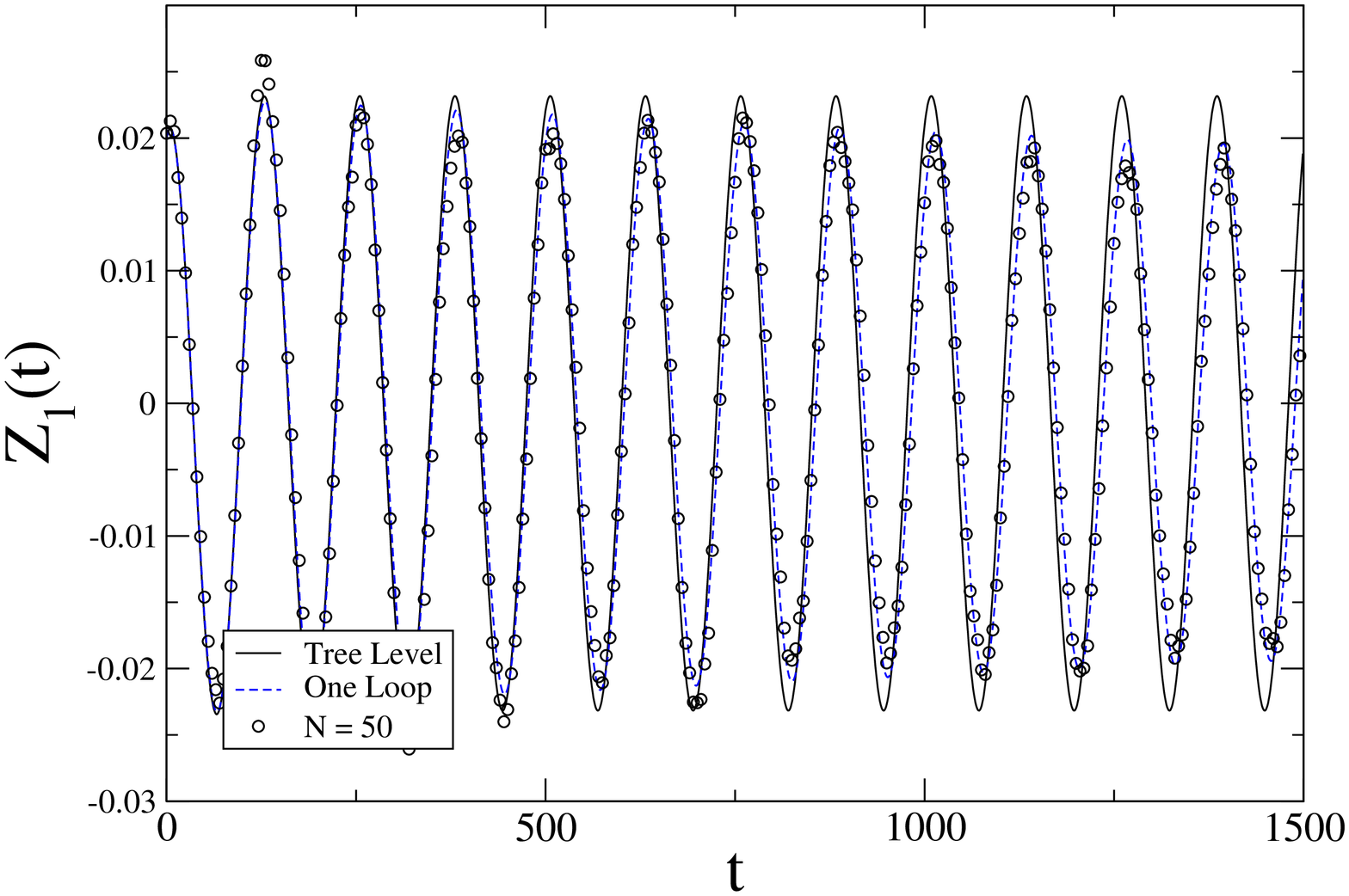}}
	\scalebox{0.25}{\includegraphics{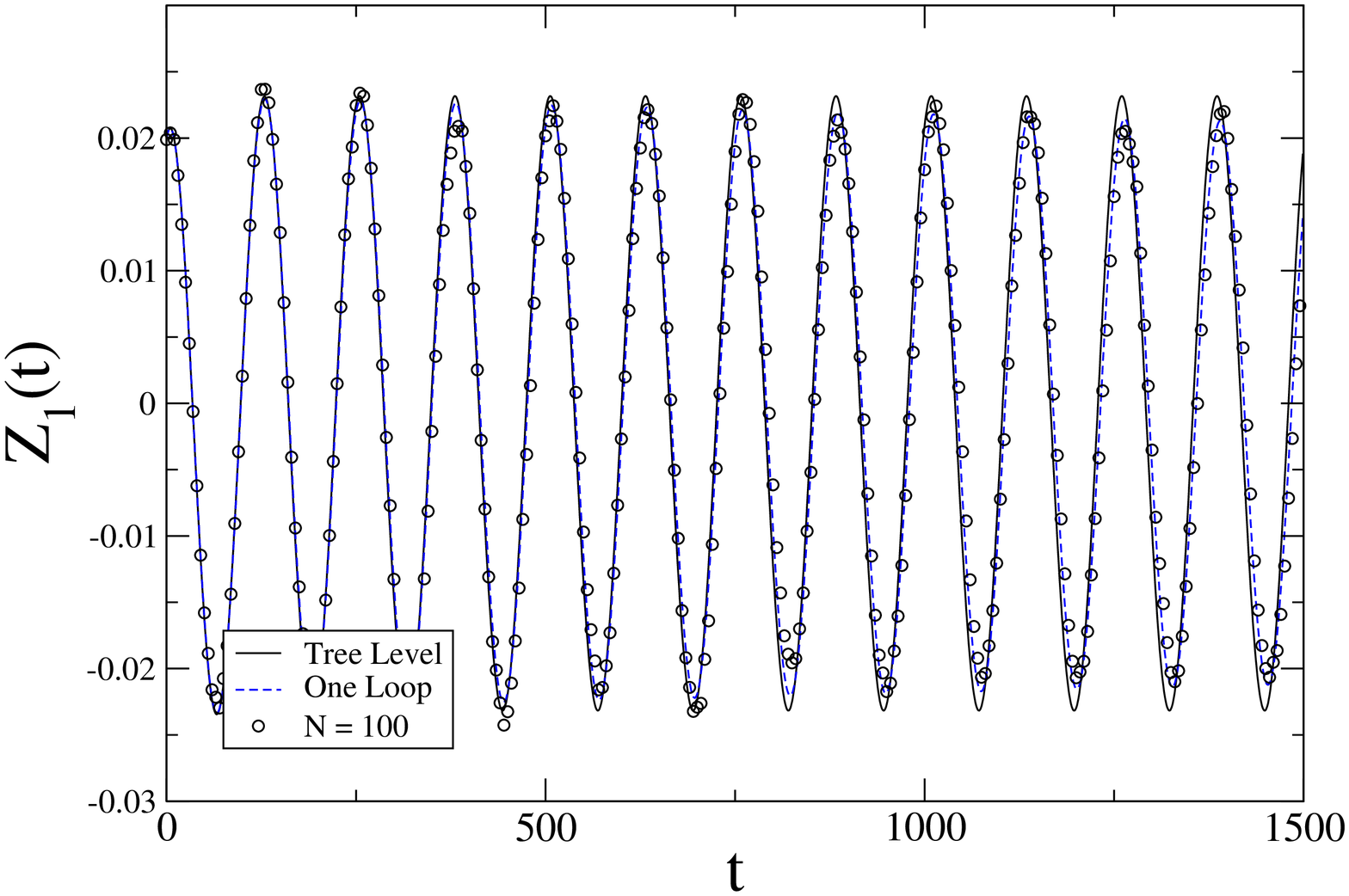}}
	\scalebox{0.25}{\includegraphics{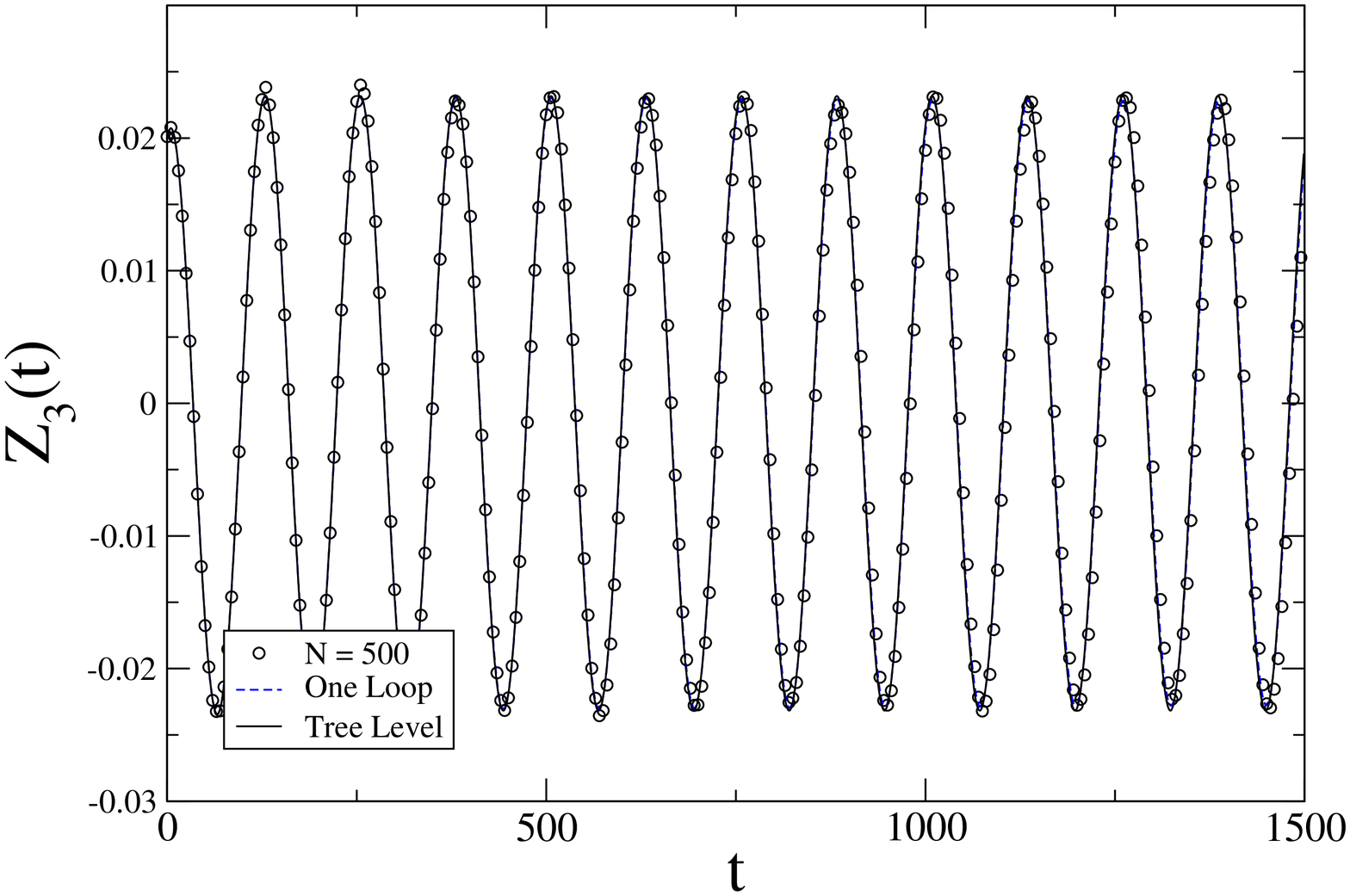}}
	\caption{As the previous figures, but with $K=0.3K_c$ and $\omega_0 = 0.05$. Solid line represents tree level.  Dashed blue line represents the one loop calculation.  Circles represent the simulation data.}
	\label{fig:omega3}
\end{figure}

\begin{figure}
	\scalebox{0.25}{\includegraphics{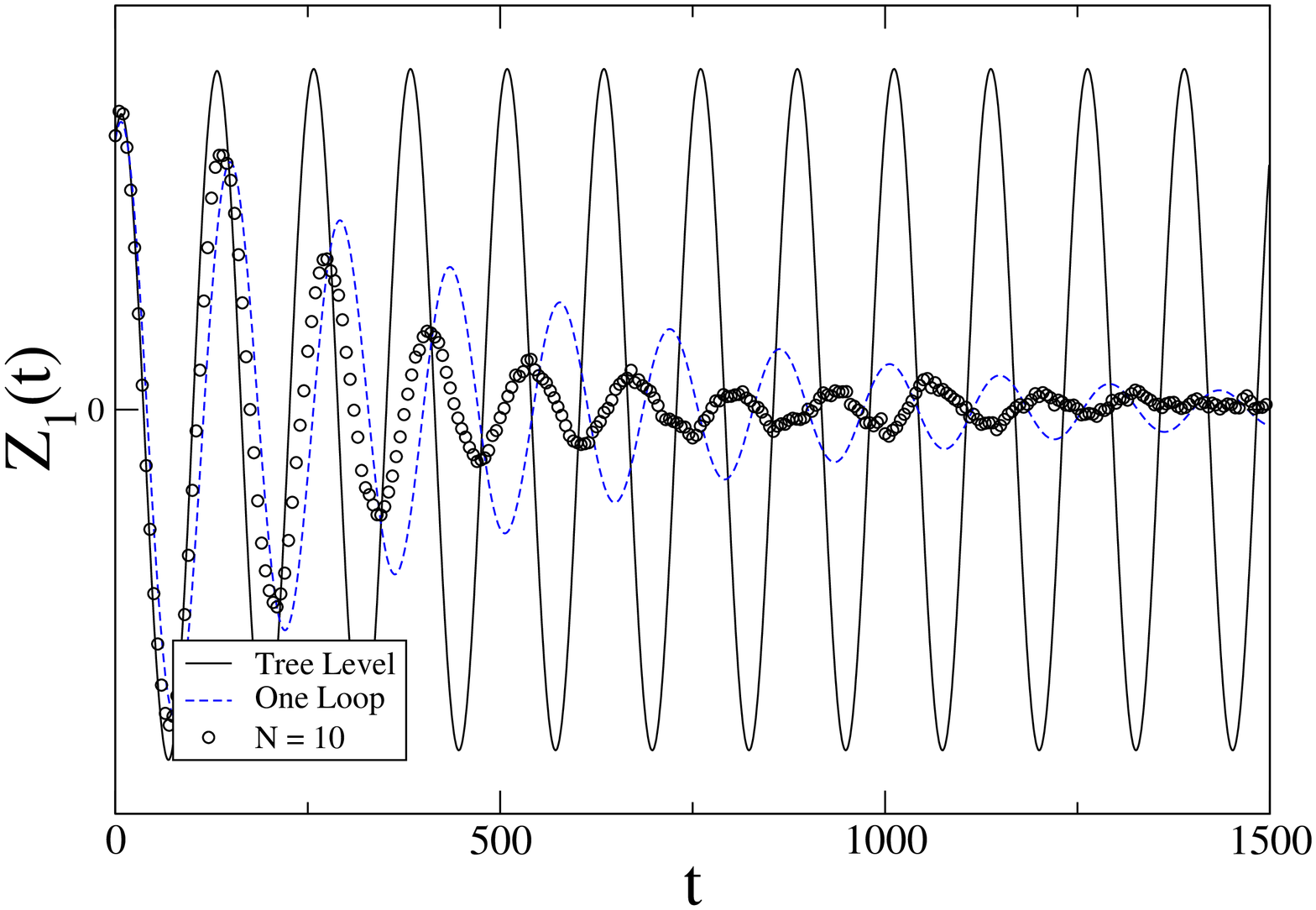}}
	\scalebox{0.25}{\includegraphics{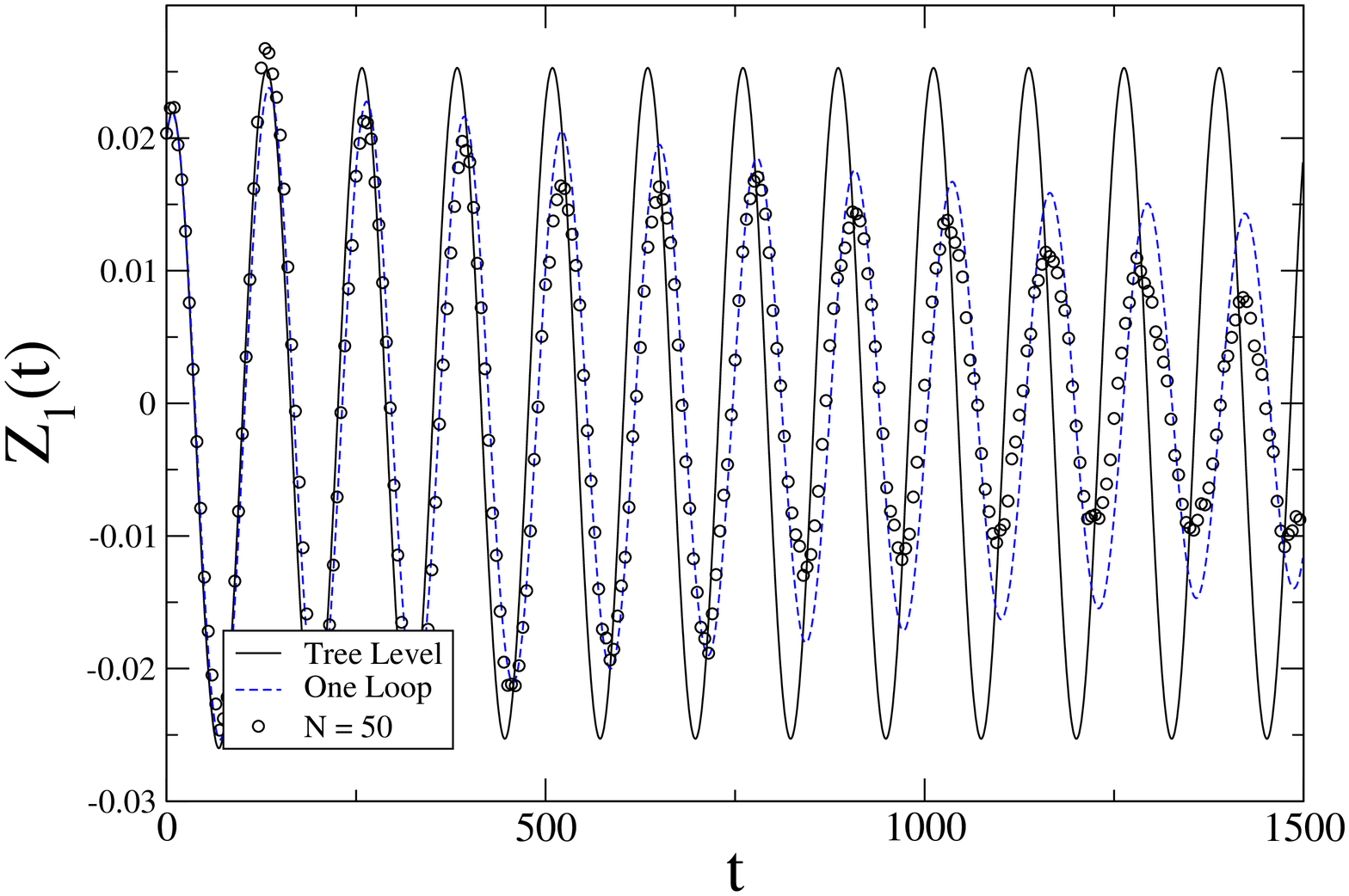}}
	\scalebox{0.25}{\includegraphics{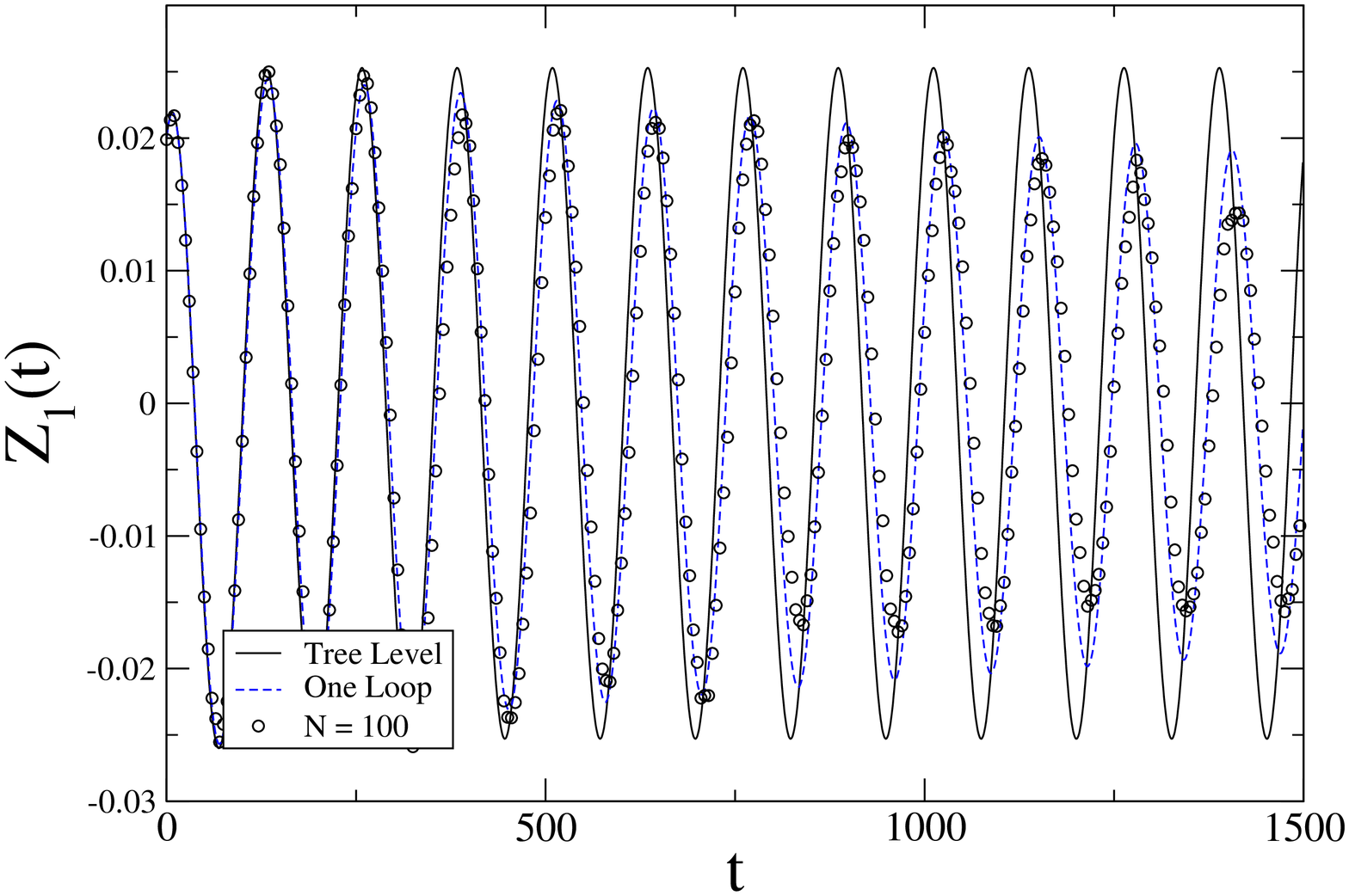}}
	\scalebox{0.25}{\includegraphics{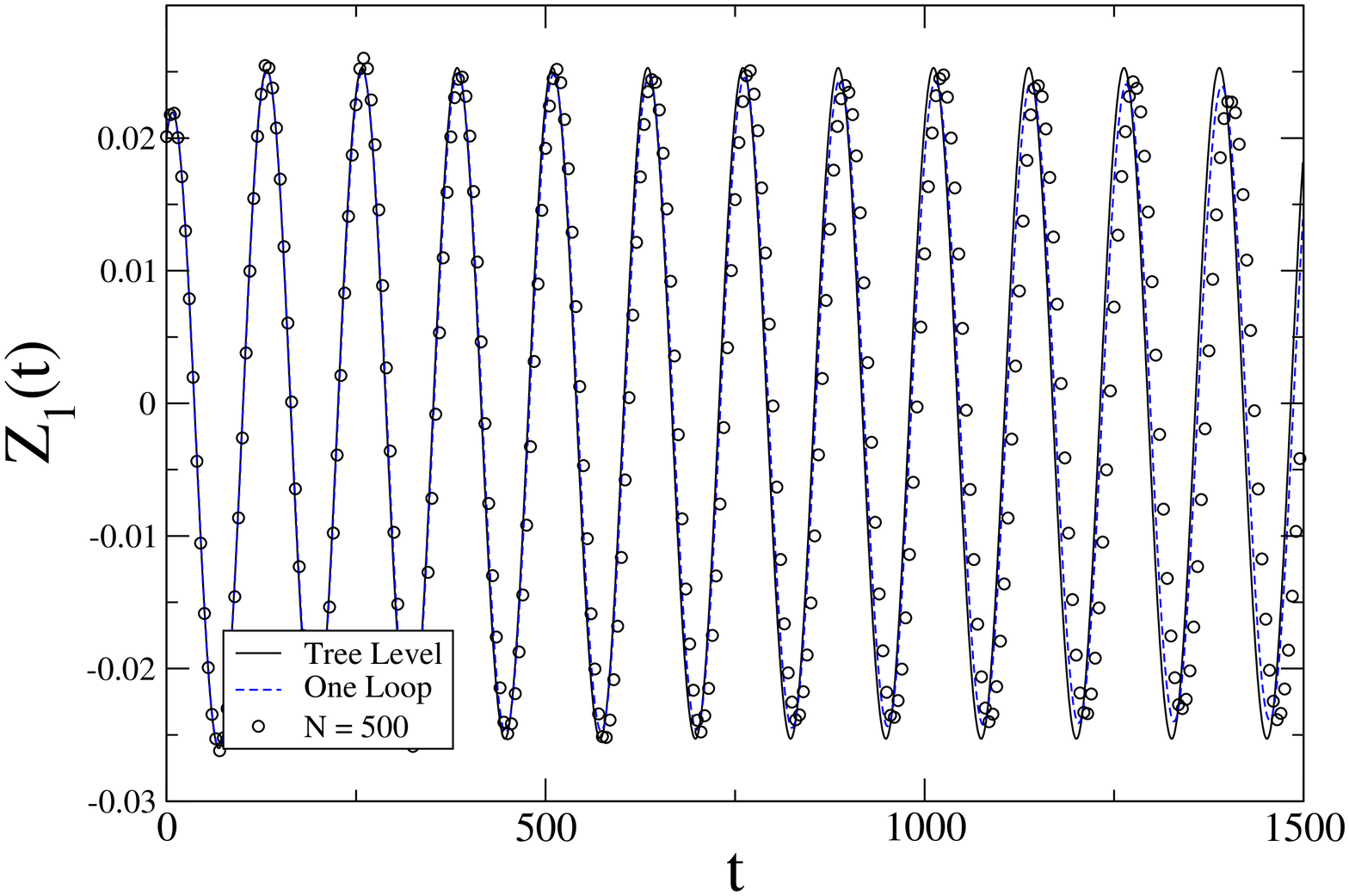}}
	\caption{As the previous figures, but with $K=0.5K_c$ and $\omega_0 = 0.05$.  Symbols as in Figure~\ref{fig:omega3}. }
	\label{fig:omega5}
\end{figure}

In Figures~\ref{fig:omega3} and \ref{fig:omega5}, we plot the time evolution of $Z_1(t)$ given that the favored oscillator has a driving frequency of $\omega_0 = 0.05$.  Note first that $Z_1(t)$ approaches the tree level calculation as $N\rightarrow \infty$.  The amplitude of the oscillation also shows the same deviation as the $\omega_0 = 0$ data, namely that of a slight initial overshoot of the one loop prediction followed by an undershoot.  In addition to this, we can see an increasing frequency shift as $N\rightarrow0$. The data, prediction, and mean field results eventually become out of phase.  For intermediate values of $N$, one can see that the one loop correction follows this shift, while for $N=10$, the mean field, data, and one loop results each have a different phase.

\begin{figure}
	\scalebox{0.25}{\includegraphics{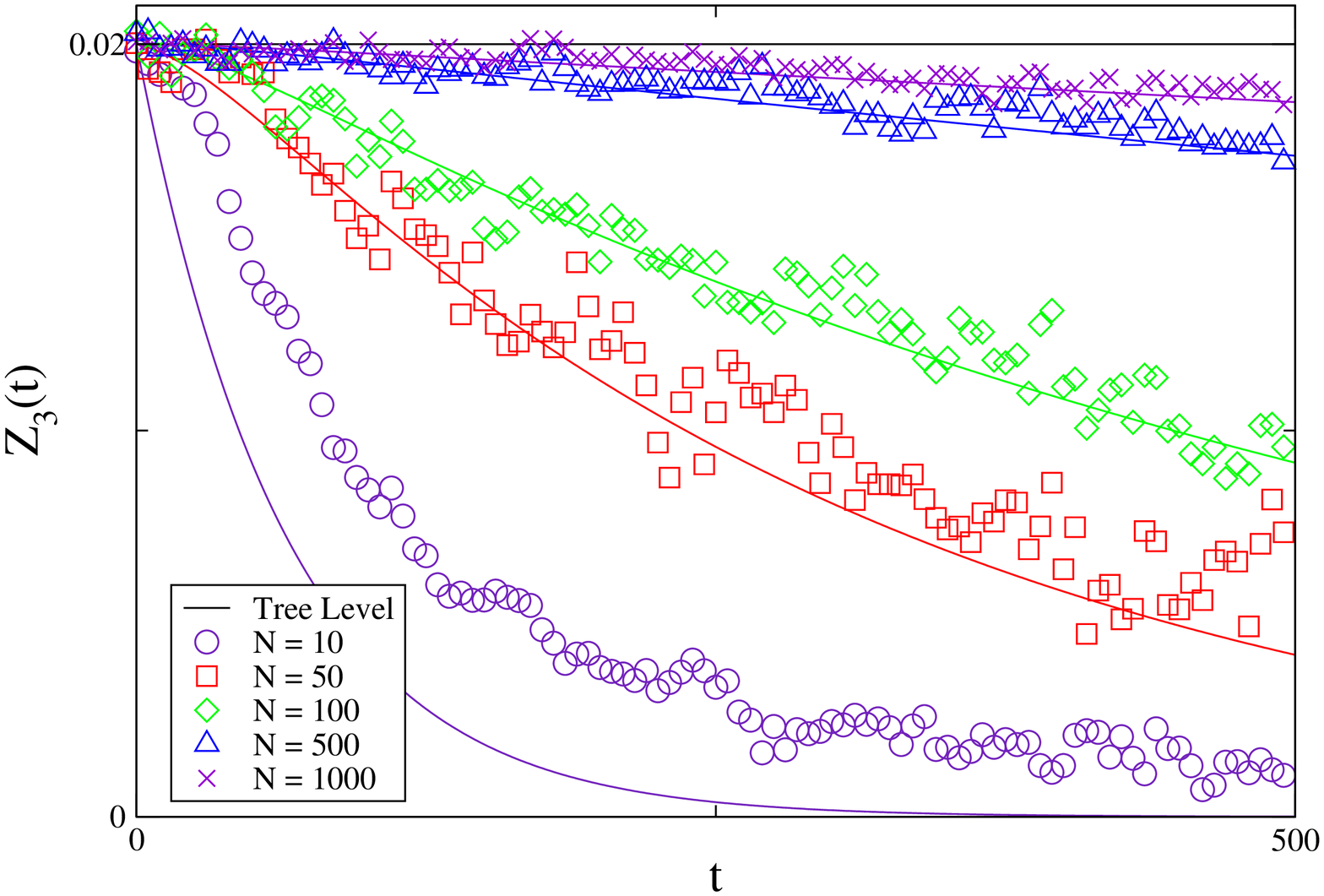}}
	\scalebox{0.25}{\includegraphics{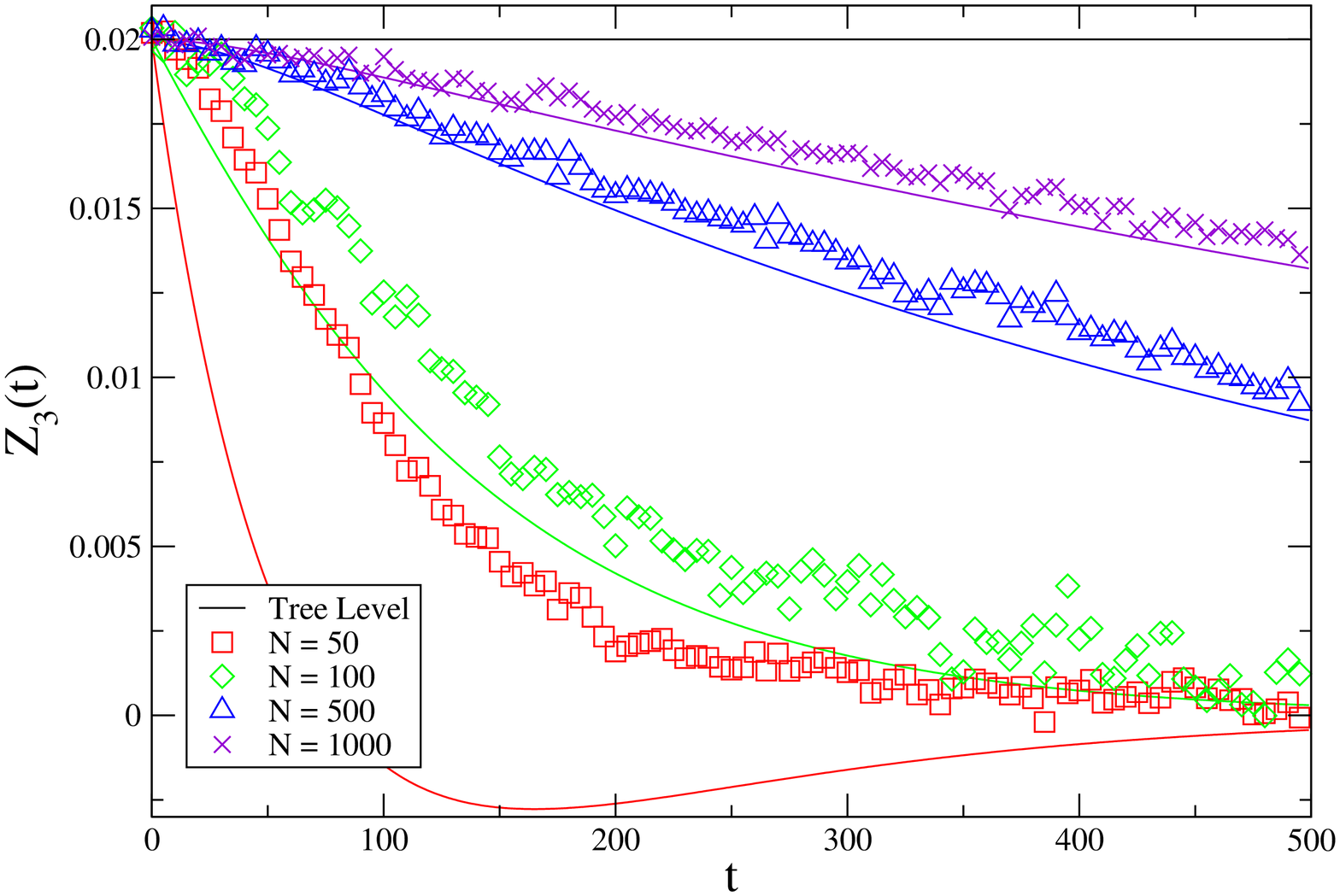}}
	\caption{$\delta Z_3(t)$ vs. $t$ for $K=0.3K_c$ (top) and $K=0.5K_c$ (bottom). Symbols as in Figure~\ref{fig:k03}.}
	\label{fig:n3}
\end{figure}

In the case of $n>2$ it is easier to write down a complete analytical solution for the time evolution. With $\omega_0 = 0$, $\theta_0 = 0$ we have, 
\begin{eqnarray}
	 Z_n(t) &=& \frac{1}{N}\left[ \left (1 + \frac{Dn^2}{ \left ( \gamma - \frac{K}{2} \right) } \right ) e^{-Dn^2 t} \right . \nonumber \\
	 &-& \left .  \frac{Dn^2}{ \left ( \gamma - \frac{K}{2} \right) }e^{-\left (\gamma - \frac{K}{2}\right)t +Dn^2 t} \right ]
\end{eqnarray}
This is compared with a simulation result in Figure~\ref{fig:n3}.  We see the same general trends as the previous graphs.  At large $N$, the simulation follows the prediction quite well.  For small $N$, the simulation seems consistently higher than the one loop prediction with $K=0.3K_c$.  For $K=0.5K_c$, the prediction is again sufficiently singular that we have not plotted $N=10$.   The deviation is already apparent for $N=50$.

\section{Discussion}

Using techniques from field theory, we have produced a theory which captures the fluctuations and correlations of the Kuramoto model of coupled oscillators.  Although we have used the Kuramoto model as an example system, the methodology is readily extendible to other systems of coupled oscillators, even those which are not interacting via all-to-all couplings.  Moreover, the methodology can be readily applied to any system which obeys a continuity equation.  
We derive an action that describes the dynamics of the Kuramoto model.  The path integral defined by this action constitutes an ensemble average over the configurations of the system, i.e. the phases and driving frequencies of the oscillators.  Because the dynamics of the model are deterministic, this is equivalent to an ensemble average over initial phases and driving frequencies.  Using the loop expansion, we can compute moments of the oscillator density function perturbatively with the inverse system size as an expansion parameter.  However, it is important to point out that the loop expansion is equivalent to an expansion in $1/N$ only because of the all-to-all coupling.  A local coupling will produce fluctuations which do not vanish in the thermodynamic  limit.

Our previous work in this direction developed a moment hierarchy analogous to the BBGKY hierarchy in plasma physics.  This paper fully encompasses that earlier work.  The equations of motion for the multi-oscillator density functions derivable from the action are in fact the equations of that BBGKY hierarchy.  The calculation in Ref.~\cite{hildebrand} is in the present context the tree level calculation of the 2-point correlation function, given by the Feynman graph in Figure~\ref{fig:connectedCorrelation}a.
With the BBGKY hierarchy, the calculational approach involves arbitrary truncation at some order, with no \emph{a priori} knowledge of how this approximation is related to the system size, $N$.  Here we show that this approximation is entirely equivalent to the loop expansion approximation.  Truncating the hierarchy at the $n$th moment is equivalent to truncating the loop expansion at the $(n-l)$th loop for the $l$th moment.  The one loop calculation is performed in the BBGKY context by considering the linear response in the presence of the 2-point correlation function.  This would produce a more roundabout manner of arriving at our one loop linear response.  One should also compare our one loop calculation with the Direct-Interaction-Approximation of fluid dynamics;  the path integral approach in that context is  the Martin-Siggia-Rose formalism \cite{martin}.  Another possible equivalent means of approaching this problem is through the Ito Calculus, treating the density as a stochastic variable and developing a stochastic differential equation for it.

An important aspect of our theory is that it is directly related to a Markov process derivable from the Kuramoto equation.  One can employ the standard Doi-Peliti method for deriving an action from a Markov process to arrive at the same theory.  Although the Kuramoto model is deterministic, the probability distribution evolves in a manner indistinguishable from a fundamentally random process.  The stochasticity of the effective Markov process is due to the distribution of phases and driving frequencies.  In other words, it is a statement about information available to us about the state of the system.  The incoherent state is a state of high entropy.  The single oscillator perturbation is one in which we have gained a small amount of information about the system and we ask a question concerning our knowledge about future states.  In the mean field limit for the single oscillator perturbation, we always know where to find the perturbed oscillator given a prescription of its initial state.  In the finite case, our ignorance of the positions and driving frequencies of the other oscillators makes  a determination of its future location difficult.  Eventually, we lose all ability to locate the perturbed oscillator as it interacts with the ``heat bath" of the population.
Furthermore, this result should be time reversal invariant.  Just as we have no way of determining with accuracy where to find the oscillator in the future, likewise we have no means of determining  where it has been at some time in the past.  To prove this statement in the context of our theory would require an analysis of the ``time reversed" theory, obtained essentially by switching the roles of $\tilde{\varphi}$ and $\varphi$.
The relevant propagator for this time reversed theory will be the solution of the linearization of the adjoint of the mean field equation.  Accordingly, this adjoint theory will have  loop corrections which will damp the time reversed propagator as well.


It is important to point out that our formulation accounts for the local stability of the incoherent state $\rho(\theta, \omega, t) = g(\omega)/2\pi$ to linear perturbations along with demonstrating the order parameters $Z_n$ approach zero.  In mean field theory, there is the possibility of quasiperiodic oscillations so that $Z_n = 0$, i.e. the modes de-phase, while the incoherent state is marginally stable and information of the initial state is retained.  Our work shows that in a finite size system, this is does not happen; the incoherent state is linearly stable.


We have considered exclusively the case of fluctuations about the incoherent state below the critical point.  Above criticality, a fraction of  the population synchronizes.  In this case, to analyze the fluctuations one may need to employ a ``low temperature" expansion in contrast  to our ``high temperature" treatment.  In essence, one separates the populations into locked and unlocked oscillators and derives a perturbation expansion from the locked action.   At criticality,  each term in the loop expansion diverges.  This is an indication that  fluctuations at all scales become relevant near the transition and thus a renormalization group approach is suggested.  Our formalism provides a natural basis for this approach.

In summary, we have provided a method for deriving the statistics of theories defined via  a Klimontovich, or continuity, equation for  a number density. This method produces a consistent means for approximating arbitrary multi-point functions.  In the case of all-to-all coupling, this approximation becomes a system size expansion.  We have demonstrated further that the system size corrections are sufficient to render the incoherent state of the Kuramoto model stable to perturbations.

\appendix









\acknowledgments{This research was supported by the Intramural Research Program of NIH/NIDDK.}

\begin{widetext}
\section{One Loop Calculation of the Propagator}
\label{app:oneloop}
The loop correction $\Gamma_{1a}$ applied to $P$ is given by
\begin{eqnarray}
	\lefteqn{\int_0^{2\pi} d\phi \int_{-\infty}^{\infty}d\nu \int_{t'}^{t} dt'\Gamma_{1a}(\theta, \omega; \phi, \nu; t - t'') P(\phi, \nu,t'';\theta', \omega'; t')=} \\
	&=&- \frac{K^2}{N} \int d\theta_1 d\omega_1 d\theta'_1 d\omega'_1 d\theta'_2 d\omega'_2 dt_1 \nonumber \\
	&& \frac{\partial}{\partial \theta} \left [ f(\theta'_2 - \theta) \left \{ P_0(\theta'_2, \omega'_2, t; \theta'_1, \omega'_1, t_1 )P_0(\theta, \omega, t; \theta_1, \omega_1, t_1 ) \right . \right . \nonumber \\
	&&+\left . \left . P_0(\theta'_2, \omega'_2, t; \theta_1, \omega_1, t_1 )P_0(\theta, \omega, t; \theta'_1, \omega'_1, t_1 )\right \}\right ] \nonumber \\
	&& \times  \frac{\partial}{\partial \theta_1} \left [ f(\theta'_1 - \theta_1) \left \{ \rho(\theta'_1, \omega'_1, t_1) P(\theta_1, \omega_1, t_1; \theta', \omega', t' ) + \rho(\theta_1, \omega_1, t_1) P(\theta'_1, \omega'_1, t_1; \theta', \omega', t' )\right \}\right ] \nonumber 
\end{eqnarray}
This term arises from one vertex with a single incoming line and two outgoing lines and one vertex with two incoming lines and a single outgoing line, hence the product of two tree level propagators, $P_0$.  We can represent $\Gamma_{1a}$ in Fourier/Laplace space as
\begin{eqnarray}
	\lefteqn{\Gamma_{1a}(n,\omega, \nu, t - t')=} \\
	&& \frac{K^2}{N} \int d\omega'_1 d\omega'_2 (2 \pi)^2 \sum_m f(m) \nonumber \\
	&& \left [ n(m+n)f(-m) g(\omega'_1) P^0(-m,\omega'_2, t; \omega'_1, t')P^0(m+n, \omega, t; \nu, t') \right . \nonumber \\
	&& (-nm)f(m) g(\omega'_1) P^0(-m,\omega'_2, t; \omega'_1, t')P^0(m+n,\omega, t; \nu, t') \nonumber \\
	&& (-mn)f(m+n) g(\omega'_1) P^0(-m,\omega'_2, t; \nu, t')P^0(m+n,\omega, t; \omega'_1, t') \nonumber \\
	&&\left . n(m+n)f(-n-m) g(\omega'_1) P^0(-m,\omega'_2, t; \nu, t')P^0(m+n,\omega, t; \omega'_1, t') \right ] \nonumber 
\end{eqnarray}
If $f(\theta)$ is odd then $f(m) = -f(-m)$.  Therefore,
\begin{eqnarray}
	\lefteqn{\Gamma_{1a}(n,\omega, \nu, t - t')=} \\
	&& \frac{K^2}{N} \int d\omega'_1 d\omega'_2 (2 \pi)^2 \sum_m f(m) \nonumber \\
	&& \left [ n(2m+n)f(-m) g(\omega'_1) P^0(-m, \omega'_2, t; \omega'_1, t')P^0(m+n,\omega, t; \nu, t') \right . \nonumber \\
	&&\left . n(2m+n)f(-n-m) g(\omega'_1) P^0(-m,\omega'_2, t; \nu, t')P^0(m+n,\omega, t; \omega'_1, t') \right ] \nonumber 
\end{eqnarray}
In this form it is easy to see the different channels which appear in the correction.

Evaluating this expression using the tree level propagator (\ref{eq:proptree}) gives us
\begin{eqnarray}
	\lefteqn{\tilde{\Gamma}_{1a}(n,\omega, \nu, s)=} \\
	&& \frac{K^2}{N}  (2 \pi)^2 \sum_m f(m) \nonumber \\
	&& \left [ n(2m+n)f(-m)  \frac{1}{2\pi }\frac{1}{imKf(m)}\left .{\rm Res}\left ( \frac{1}{\Lambda_{-m}(s_1)}  \right )\right|_{s_1=s_n} N\tilde{P}^0(m+n,\omega; \nu, s-s_n) \right . \nonumber \\
	&& n(2m+n)f(-m-n)  \frac{1}{2\pi } \left .{\rm Res}\left ( \frac{1}{\Lambda_{-m}(s_1)}  \right )\right|_{s_1=s_n}  \frac{1 }{\left ( s_n - im \nu \right )}\frac{1}{2\pi } \frac{g(\omega)}{\left (s - s_n +i(m+n)\omega \right)} \frac{1}{\Lambda_{m+n}(s-s_n)} \nonumber \\
	&&\left . n(2m+n)f(-m-n)  \frac{1}{2\pi } \frac{1}{\Lambda_{-m}(i m \nu)} \frac{1}{2\pi } \frac{g(\omega)}{\left (s - im \nu +i(m+n)\omega \right)} \frac{1}{\Lambda_{m+n}(s-im \nu)} \right ]\nonumber \\
\end{eqnarray}

The diagram $\Gamma_{1c}$ is given by
\begin{eqnarray}
\lefteqn{\int_0^{2\pi} d\phi \int_{-\infty}^{\infty}d\eta \int_0^{t} dt'\Gamma_{1c}(n,\theta, \omega; \phi, \nu; t - t') P(\phi, \nu;\theta', \omega'; t')=} \\
	&& 2 N^2 \frac{K^3}{(2\pi)^2} \frac{\partial}{\partial \theta} \int d\theta'_3 d\omega'_3 \prod_{i=1}^{2} d\theta'_i d\omega'_i d\theta_i d\omega_i dt_i \nonumber \\
	&&\left [ f(\theta'_3 - \theta) \frac{\partial}{\partial \theta_1} f(\theta'_1 - \theta_1) g(\omega_1) g(\omega'_1) \right . \nonumber \\
	&&\left \{ P_0(\theta'_3, \omega'_3, t; \theta_2 , \omega_2, t_2) P_0(\theta, \omega, t; \theta_1 , \omega_1, t_1)  \right . \nonumber \\
	&&+\left . P_0(\theta'_3, \omega'_3, t; \theta_1 , \omega_1, t_1) P_0(\theta, \omega, t; \theta_2 , \omega_2, t_2) \right \} \nonumber \\
	&&\frac{\partial}{\partial \theta_2} f(\theta'_2 - \theta_2) \left \{ P_0(\theta'_2, \omega'_2, t_2; \theta'_1 , \omega'_1, t_1) P(\theta_2, \omega_2, t_2; \theta' , \omega', t') \right . \nonumber \\
	&&+\left . \left .  P(\theta'_2, \omega'_2, t_2; \theta' , \omega', t') P_0(\theta_2, \omega_2, t_2; \theta'_1 , \omega'_1, t_1) \right \} \right ]
\end{eqnarray}
In Fourier space, we can write:
\begin{eqnarray}
	\Gamma_{1c}(n,\omega, \nu, t - t') = 2 N^2 K^3 (2\pi)^3 \sum_{m}\int d\omega'_3   d\omega_1 d \omega'_1  dt_1   g(\omega_1) g(\omega'_1)  \nonumber \\
	 \left [ \int d\omega'_2   (imn)(m+n) f(-m-n)f(m) f(-m) P(n+m,\omega'_3, t; \nu, t') P(-m,\omega, t; \omega_1, t_1) P(m,\omega'_2, t'; \omega'_1, t_1) \right . \nonumber \\
	+ \int d\omega'_2   inm(m+n) f(m)f(m) f(-m) P(-m,\omega'_3, t; \omega_1, t_1) P(n+m,\omega, t; \nu, t') P(m,\omega'_2, t'; \omega'_1, t_1) \nonumber \\
	 + \int d\omega_2   inm(m+n) f(-n-m)f(m) f(-n) P(n+m,\omega'_3, t; \omega_2, t') P(-m,\omega, t; \omega_1, t_1) P(m,\omega_2, t'; \omega'_1, t_1) \nonumber \\
	+ \left . \int d\omega_2   inm(m+n) f(m)f(m) f(-n) P(-m,\omega'_3, t; \omega_1, t_1) P(n+m,\omega, t; \omega_2, t') P(m,\omega_2, t'; \omega'_1, t_1) \right ]
\end{eqnarray}
and taking the Laplace transform:
\begin{eqnarray}
	\tilde{\Gamma}_{1c}(n,\omega, \nu, s) = \left( \frac{1}{2\pi i}\right)2 N^2 K^3 (2\pi)^3 \sum_{m}\int d\omega'_3   d\omega_1 d \omega'_1  ds_1   g(\omega_1) g(\omega'_1)  \nonumber \\
	\left [ \int d\omega'_2   (imn)(m+n) f(-m-n)f(m) f(-m) \tilde{P}(n+m,\omega'_3; \nu, s - s_1) P(-m,\omega; \omega_1, s_1) P(m)\omega'_2; \omega'_1, -s_1)\right .  \nonumber \\
	+ \int d\omega'_2   inm(m+n) f(m)f(m) f(-m) P(-m,\omega'_3; \omega_1, s_1) P(n+m,\omega; \nu, s-s_1) P(m,\omega'_2; \omega'_1, -s_1) \nonumber \\
	+ \int d\omega_2   inm(m+n) f(-n-m)f(m) f(-n) P(n+m,\omega'_3; \omega_2, s-s_1) P(-m,\omega; \omega_1, s_1) P(m,\omega_2; \omega'_1, -s_1) \nonumber \\
	+ \left . \int d\omega_2   inm(m+n) f(m)f(m) f(-n)P(-m,\omega'_3; \omega_1, s_1) P(n+m,\omega; \omega_2, s-s_1) P(m,\omega_2; \omega'_1, -s_1) \right ]
\end{eqnarray}
where the contour for $s_1$ lies between 0 and $0<s<-{\rm Re}(s_n)$.
Performing the integrals, we have:
\begin{eqnarray}
	\tilde{\Gamma}_{1c}(n,\omega, \nu, s) &=& \frac{2}{N} K^3 \left (\frac{1}{2\pi i} \right )\int ds_1 \sum_{m}  (imn)(m+n)\nonumber \\
	&&\left [ f(-m-n) f(m) f(-m)\frac{1}{\Lambda_{n+m}(s-s_1)} \frac{1}{s - s_1 + i(m+n) \nu}\frac{g(\omega)}{s_1 - im\omega}\frac{1}{\Lambda_{-m}(s_1)}  \right . \nonumber \\
	&&\times \left (\frac{-1}{imKf(-m)}\right ) \left (\frac{1}{\Lambda_m(-s_1)} - 1 \right ) \nonumber \\
	&+& f(m)f(m) f(-m) \left (\frac{1}{imKf(m)}\right ) \left (\frac{1}{\Lambda_{-m}(s_1)} - 1 \right )(2\pi N)P(n+m,\omega; \nu, s-s_1) \nonumber \\
	&& \times  \left (\frac{-1}{imKf(-m)}\right ) \left (\frac{1}{\Lambda_m(-s_1)} - 1 \right ) \nonumber \\
	&+& \int d\omega_2   f(-n-m)f(m) f(-n) \frac{1}{\Lambda_{n+m}(s-s_1)} \frac{1}{s - s_1 + i(m+n)\omega_2} \frac{g(\omega)}{s_1 - im\omega}\frac{1}{\Lambda_{-m}(s_1)} \nonumber \\
	&& \times \frac{g(\omega_2)}{-s_1 + im\omega_2}\frac{1}{\Lambda_{m}(-s_1)}  \nonumber \\
	&+&  	\int d\omega_2   f(m)f(m) f(-n)  \left (\frac{1}{imKf(m)}\right ) \left (\frac{1}{\Lambda_{-m}(s_1)} - 1 \right )(2\pi N)P(n+m,\omega; \omega_2, s-s_1) \nonumber \\
	&& \left .\times  \frac{g(\omega_2)}{-s_1 + im\omega_2}\frac{1}{\Lambda_{m}(-s_1)}  \right ] \nonumber \\
\end{eqnarray}

Finally, the diagram $\Gamma_{1d}$ is given by
\begin{eqnarray}
\lefteqn{\int_0^{2\pi} d\phi \int_{-\infty}^{\infty}d\nu \int_0^{t} dt'\Gamma_{1d}(\theta, \omega; \phi, \nu; t - t') P(\phi, \nu;\theta', \omega'; t')=} \\
	&& -  N^2 \frac{K^2}{(2\pi)^2} \frac{\partial}{\partial \theta} \int d\theta'_3 d\omega'_3 dt_2\prod_{i=1}^{2} d\theta'_i d\omega'_i d\theta_i d\omega_i  \nonumber \\
	&&\left [ f(\theta'_3 - \theta) g(\omega_1) g(\omega'_1) \right . \nonumber \\
	&&\left \{ P_0(\theta'_3, \omega'_3, t; \theta_2 , \omega_2, t_2) P_0(\theta, \omega, t; \theta_1 , \omega_1, t_0)  \right . \nonumber \\
	&&+\left . P_0(\theta'_3, \omega'_3, t; \theta_1 , \omega_1, t_0) P_0(\theta, \omega, t; \theta_2 , \omega_2, t_2) \right \} \nonumber \\
	&&\frac{\partial}{\partial \theta_2} f(\theta'_2 - \theta_2) \left \{ P_0(\theta'_2, \omega'_2, t_2; \theta'_1 , \omega'_1, t_0) P(\theta_2, \omega_2, t_2; \theta' , \omega', t') \right . \nonumber \\
	&&+\left . \left .  P(\theta'_2, \omega'_2, t_2; \theta' , \omega', t') P_0(\theta_2, \omega_2, t_2; \theta'_1 , \omega'_1, t_0) \right \} \right ]
\end{eqnarray}

Using the fact that $\int d\omega' d\theta' P(\theta, \omega, t; \theta', \omega', t') g(\omega') = g(\omega)/N$ and $\int d\theta f(\theta) = 0$ we have
\begin{eqnarray}
\lefteqn{\int_0^{2\pi} d\phi \int_{-\infty}^{\infty}d\nu \int_0^{t} dt'\Gamma_{1d}(\theta, \omega; \phi, \nu; t - t') P(\phi, \nu;\theta', \omega'; t')=} \\
	&& -   \frac{K^2}{(2\pi)^2} \frac{\partial}{\partial \theta} \int d\theta'_3 d\omega'_3 dt_2\prod_{i=1}^{2} d\theta'_i d\omega'_i d\theta_i d\omega_i  \nonumber \\
	&&\left [ f(\theta'_3 - \theta) g(\omega_1) g(\omega'_1)  P_0(\theta'_3, \omega'_3, t; \theta_2 , \omega_2, t_2)   \right .\nonumber \\
	&&\frac{\partial}{\partial \theta_2} f(\theta'_2 - \theta_2)  \left .  P(\theta'_2, \omega'_2, t_2; \theta' , \omega', t')  \right ]
\end{eqnarray}

In Fourier-Laplace we have
\begin{eqnarray}
	\tilde{\Gamma}_{1d}(n; \omega, \nu; s ) =   \frac{1}{N} in K f(-n) g(\omega)  \left ( \frac{1}{\Lambda_n(s)} -1 \right ) 
\end{eqnarray}
\end{widetext}

\section{ $f(\theta) = \sin \theta$ and $g(\omega)$ Lorentz  }
\label{app:lorentz}
	In order to both simplify the correction and to provide a concrete example, we will specialize to the case that $g(\omega)$ is a Lorentz distribution and $f(\theta) = \sin \theta$.  $f(\theta) = \sin \theta$ is the traditional coupling for the Kuramoto model (and has the advantage of being bounded in Fourier space so that we avoid ``ultraviolet" singularities) and using a Lorentz frequency distribution yields analytical results.
	
The Lorentz frequency distribution is given by
\begin{equation}
	g(\omega) = \frac{1}{\pi} \frac{\gamma}{\gamma^2 + \omega^2}
\end{equation}
From this we can calculate
\begin{equation}
	\Lambda_{\pm 1}(s) = \frac{s + \gamma - \frac{K}{2}}{s + \gamma}
\end{equation}
and $\Lambda_n(s) = 1$ for $n\ne \pm 1$.  The residue of the function $\Lambda_{\pm 1}(s)$ is
\begin{equation}
	 \left .{\rm Res}\left ( \frac{1}{\Lambda_{-m}(s_1)}  \right )\right|_{s_1=s_n} = \frac{K}{2}
\end{equation}
and the pole is at $s_{\pm 1} = - (\gamma - K/2)$.  This provides a critical coupling 
\begin{equation}
	K_c = 2\gamma
\end{equation}
above which the system begins to synchronize.  The incoherent state is reached when $K < K_c$, which gives $s_\pm < 0$.
From $f(\theta) = \sin \theta$
we also have $f(\pm 1) = \mp i/2$.

In this case, diagram $a$ evaluates to
\begin{widetext}
\begin{eqnarray}
	\lefteqn{\tilde{\Gamma}_{1a}(n,\omega, \nu, s)=} \\
	&& -\frac{K^2}{N}  (2 \pi)^2 \sum_{m = \pm 1} \frac{i}{2}m \nonumber \\
	&& \left [ n(2m+n)f(-m)  \frac{1}{2\pi }\frac{1}{imKf(m)}\frac{K}{2}N\tilde{P}^0(m+n,\omega; \nu, s+\gamma - \frac{K}{2}) \right . \nonumber \\
	&& n(2m+n)f(-m-n)  \frac{1}{2\pi } \frac{K}{2}  \frac{1 }{\left ( \frac{K}{2} - \gamma - im \nu \right )}\frac{1}{2\pi } \frac{g(\omega)}{\left (s +\gamma - \frac{K}{2} +i(m+n)\omega \right)} \frac{s + 2\gamma - \frac{K}{2}}{s + 2\gamma - K} \nonumber \\
	&& \left .n(2m+n)f(-m-n)  \frac{1}{2\pi } \frac{1}{\Lambda_{-m}(i m \nu)} \frac{1}{2\pi } \frac{g(\omega)}{\left (s - im \nu +i(m+n)\omega \right)} \frac{s + \gamma - i m \nu}{s + \gamma - \frac{K}{2} - im \nu} \right ] \nonumber \\
\end{eqnarray}
\end{widetext}
There is no contribution for $n=0$ which we expect from probability conservation.  The terms proportional to $n(2m+n)f(-n-m)$ will always evaluate to $0$, because $f(n\ne \pm 1) = 0$.
Since the tree level propagator contains such a term as well,  we have the simplification
\begin{eqnarray}
	&&\tilde{\Gamma}_{1a}(n,\omega, \nu, s) =   \\
	&& \frac{K^2}{N}   \sum_{m = \pm 1} \frac{1}{4} 
	 \left [ n(2m+n) \frac{\delta(\omega - \nu)}{s + \gamma - \frac{K}{2} + i(m+n)\omega}\right ]  \nonumber 
\end{eqnarray}

For diagram $c$, we see that we can immediately ignore the third term because $nmf(-n)f(-m-n)$ is always $0$.  We also see that the first term is only non-zero for $n= \pm 2$, and the last term only for $n=\pm 1$.  
After performing the $s_1$ integration, this leaves us with
\begin{widetext}
\begin{eqnarray}
	\lefteqn{\tilde{\Gamma}_{1c}(n,\omega, \nu, s) =}   \nonumber \\
	&&\frac{K^3}{2N} \left [ \frac{1}{2}\sum_{m}  n(m+n) \frac{\delta(\omega - \nu)}{s +\gamma - \frac{K}{2} + i(n+m) \omega}\frac{1}{ 2\gamma -K}\right. \nonumber \\
	&&+ \left . 	\delta_{n\pm1}  \frac{1}{s +\gamma - \frac{K}{2} + 2in \omega}
	 \frac{g(\omega)}{\gamma - \frac{K}{2} + in\omega}\left (\frac{2\gamma - \frac{K}{2}}{2 \gamma - K} \right )  \right .\nonumber \\
	 && + \left . \delta_{n\pm 2} (-g(\omega))\left (  \frac{s - \frac{n}{2} i \omega}{s - \frac{n}{2}i\omega + \gamma - \frac{K}{2}}   \frac{1}{s + \frac{n}{2} i (\nu - \omega)} \frac{\frac{n}{2}i\omega + \gamma}{\left (\gamma - \frac{K}{2} \right )^2 + \omega^2} \right . \right . \nonumber \\
	 && \left . \left . + \frac{s + \gamma - \frac{K}{2}}{s + 2\gamma - K}  \frac{1}{s + \gamma - \frac{K}{2} + \frac{n}{2} \nu}\frac{1}{2\gamma - K} \frac{1}{-\gamma + \frac{K}{2} - \frac{n}{2}i\omega} \frac{K}{2}  \right. \right . \nonumber \\
	 && \left . \left . + \frac{n}{2} \frac{K}{2} g(\omega)\frac{1}{2\gamma - K} \frac{1}{s + \gamma - \frac{K}{2} + \frac{n}{2}i\omega } \frac{1}{s + \gamma - \frac{K}{2} + \frac{n}{2}i \nu } \frac{s + 2\gamma - \frac{K}{2}}{s + 2\gamma - K}\right ) \right ]
\end{eqnarray}
\end{widetext}

$\Gamma_{1d}$ is given simply by
\begin{eqnarray}
	\tilde{\Gamma}_{1d}(\pm 1; \omega, \nu; s ) &=&  - \frac{1}{N} g(\omega)\frac{K}{2} \frac{ \frac{K}{2} }{ s + \gamma - \frac{K}{2}}  \nonumber \\
	\tilde{\Gamma}_{1d}(n \ne \pm 1; \omega, \nu; s ) &=& 0 
\end{eqnarray}

\section{Equivalent Markov Process}

\label{app:markov}

The action (\ref{eq:action}) can be derived by applying the Doi-Peliti method to a Markov process equivalent to the Kuramoto dynamics.
Consider a two dimensional lattice ${\mathcal L}$, periodic in one dimension, with lattice constants $a_{\theta}$ in the periodic direction and $a_{\omega}$ in the other.  (The radius of the eventual cylinder is $R$.)  The indices $i$ and $j$ will be used for the frequency and periodic domains, respectively.  The oscillators obey an equation of the form:
\begin{equation}
	\dot{\theta}_i =  v \left ( \vec{\theta}, \vec{\omega}  \right )
	\label{eq:basiceom}
\end{equation}
The indices on $\vec{\theta}$ and $\vec{\omega}$ run over the lattice points of the periodic and frequency variables.  The state of the system is described by the number of oscillators $n_{i,j}$ at each site.  Given this, the fraction of oscillators found on the lattice sites is governed by the following Master equation:
\begin{eqnarray}
	\frac{dP\left ( \vec{n} , t \right)}{dt} &=& \sum_{ij} \left [ - \frac{v_{i,j}}{a_{\theta}}n_{i,j} P\left ( \vec{n} , t \right) \right . \nonumber \\
	&+& \left .  \frac{v_{i,j-1}}{a_{\theta}} (n_{i,j-1}+1)P\left ( \vec{n}^{ j } , t \right)\right ]
	\label{eq:masterequation1}
\end{eqnarray}
where the indices of the vector $\vec{n}$ run over the lattice points, ${\mathcal L}$, and $\vec{n}^{ j}$ (note the superscript) is equal to $\vec{n}$ except for the $j$th and $j-1$st components.  At those points we have $n^j_{i,j} = n_{i,j} - 1$ and $n^j_{i,j-1}=n_{i,j-1} + 1$.   The first term on the RHS represents the outward flux of oscillators from the state with $n_j$ oscillators at each periodic lattice point while the second term is the inward flux due to oscillators ``hopping" from $j-1$ to $j$.  There is no flux in the other direction ($\omega$); this lattice variable simply serves to label each oscillator by its fundamental frequency.

Consider a generalization of the Kuramoto model of the form:
\begin{equation}
	\dot{\theta}_i = \omega_i + \frac{K}{N} \sum_j f\left (  \theta_j - \theta_i \right)
\end{equation}
$N$ is the total number of oscillators and we impose $f(0) = 0$.  The velocity in equation~(\ref{eq:basiceom}) now has the form:
\begin{equation}
	v_{ij} = ia_{\omega} + \frac{K}{N} \sum_{i', j'} f\left ( [j'-j]a_{\theta} \right) n_{i',j'}
\end{equation}
In the limit $a_{\omega} \rightarrow 0$ we have $ia_{\omega} = \omega$.  Similarly, $ia_{\theta} \rightarrow \theta$.  The factor of $n_{i',j}$ has been added because the sum must cover \emph{all} oscillators, and this factor describes the number at each site.  We also sum over all frequency sites $i'$.  The master equation~(\ref{eq:masterequation1}) now takes the form:
\begin{eqnarray}
	\lefteqn{\frac{dP\left ( \vec{n} , t \right)}{dt} =} \nonumber \\
	&& \sum_{ij} \left [ - \left (\frac{ia_{\omega}}{a_{\theta}} + \frac{K}{Na_{\theta}} \sum_{i', j'} f\left ( [j'-j]a_{\theta} \right)n_{i',j'} \right ) \right .  \nonumber \\ 
	&&\times n_{i,j} P\left ( \vec{n} , t \right) \nonumber \\
	&& +   \left (\frac{ia_{\omega}}{a_{\theta}} + \frac{K}{Na_{\theta}} \sum_{i', j'} f\left ( [j'-j + 1]a_{\theta} \right) n_{i',j'} \right ) \nonumber \\
	&&\left . \times (n_{i,j-1}+1)P\left ( \vec{n}^{ j } , t \right)\right ] \nonumber \\
	&=& - {\cal H}P(\vec{n}, t)
	\label{eq:masterequation}
\end{eqnarray}
The matrix ${\mathcal H}$ is the Hamiltonian.  From this point, one can develop an operator representation as in Doi-Peliti.  Using coherent states and taking the continuum and thermodynamic limits results in the action~(\ref{eq:action}), after ``shifting" the field $\tilde{\varphi}$.

\bibliography{kineticRefs.bib}

\end{document}